\documentclass[11pt,3p,review,authoryear]{elsarticle}

\journal{International Journal of Forecasting}
\biboptions{authoryear,round}

\usepackage[authoryear,round]{natbib}
\usepackage{amsmath,amssymb}
\usepackage{mathtools}
\usepackage{graphicx}%
\usepackage{xcolor}
\usepackage{geometry}
\usepackage{floatrow}
\usepackage{float}
\usepackage{setspace}
\usepackage{booktabs}
\usepackage{pdflscape} 
\usepackage{tikz}
\usetikzlibrary{positioning}
\usetikzlibrary{arrows.meta, positioning}
\usepackage{eurosym}
\usepackage{amsfonts}
\usepackage{amsthm}
\usepackage{hyperref}
\usepackage{accents}
\usepackage{float,rotating,subfigure}
\usepackage{longtable,lscape}
\usepackage{blindtext,pdflscape}
\usepackage{multirow}
\usepackage{rotating}
\usepackage{threeparttable}
\usepackage{multirow} 
\usepackage{setspace}
\usepackage{amsmath}%
\usepackage{changepage}
\usepackage{soul}
\usepackage{url}
\usepackage{bbm}
\providecommand{\U}[1]{\protect\rule{.1in}{.1in}}

\hypersetup{colorlinks=true,
	linkcolor=blue,
	citecolor=blue,
	urlcolor=blue}
\numberwithin{equation}{section}
\counterwithin{figure}{section}
\counterwithin{table}{section}
\usepackage[shortlabels]{enumitem}

\newcommand{\RNum}[1]{\uppercase\expandafter{\romannumeral #1\relax}}
\newcommand{\rNum}[1]{\lowercase\expandafter{\romannumeral #1\relax}}
\DeclarePairedDelimiter\abs{\lvert}{\rvert}%
\DeclarePairedDelimiter\norm{\lVert}{\rVert}%
\makeatletter
\let\oldabs\abs
\def\abs{\@ifstar{\oldabs}{\oldabs*}}
\let\oldnorm\norm
\def\norm{\@ifstar{\oldnorm}{\oldnorm*}}
\makeatother

\hypersetup{
	colorlinks=true,
	linkcolor=blue,
	citecolor=blue,
	urlcolor=blue
}

\newcommand{\R}{\mathbb{R}}

\newcommand{\N}{\mathbb{N}}

\begin{document}

\begin{frontmatter}

\title{A Comparison of High-Dimensional Variable Selection Procedures for Electricity Spot Price Forecasting}

\author[aau]{Charisios Grivas}

\author[aau,norlys]{Mikkel Mandrup\corref{cor}}
\ead{mmni@math.aau.dk}

\author[aau]{Orimar Sauri}

\cortext[cor]{Corresponding author}
\address[aau]{Aalborg University}
\address[norlys]{Norlys Energy Trading A/S}


\begin{abstract}
The paper considers the problem of variable selection for forecasting electricity spot prices. High-dimensional methods such as LASSO and Elastic Net are widely used for this purpose, and while they exhibit strong predictive performance, their tendency to select over-parameterized models raises questions about interpretability. We evaluate the performance of six variable selection procedures, including the recently proposed Boosting Multiple Testing (BMT) method, using an extensive dataset from six regional electricity markets. We assess their performance in terms of both out-of-sample forecasting accuracy and model parsimony.
We find that, although LASSO and Elastic Net achieve similar accuracy and outperform most screening alternatives, BMT matches their forecasting performance while using less than one-tenth as many variables.  Our results reveal that BMT offers researchers and practitioners a substantially more interpretable and computationally efficient alternative to shrinkage methods, without any loss of forecasting accuracy. These findings suggest that the over-parameterization typically associated with regularization methods is not a necessary price for predictive accuracy in electricity price forecasting.
\end{abstract}

\begin{keyword}
Electricity price forecasting \sep Variable selection\sep LASSO \sep  Boosting \sep Multiple testing
\end{keyword}

\end{frontmatter}

\section{Introduction}
Accurate forecasting of electricity spot prices is vital for market participants, system operators and policymakers alike, informing decisions ranging from bidding strategies and risk management to the design of regulatory interventions. Electricity markets are characterised by pronounced volatility, strong seasonality, and price dynamics that respond to a wide array of fundamentals - demand, fuel costs, weather conditions and renewable generation, and cross-border flows  \citep{WERON20141030, weron2006modeling}. As a result, electricity price forecasting (EPF) models routinely draw on large sets of candidate predictors, often exceeding the number of available observations once lagged values and interactions are included. This has made EPF a natural setting for the application of high-dimensional variable selection techniques, yet the range of methods actually employed in this literature remains comparatively narrow.
	
	To date, penalised-regression approaches such as the Least Absolute Shrinkage and Selection Operator (LASSO) \citep{lasso} and the Elastic Net \citep{ElasticNet} have dominated applied work in EPF, owing to their computational tractability and strong predictive performance. \citet{ziel2015efficient} were among the first to apply LASSO to sparsify large parameter sets when estimating hourly electricity prices on the European Power Exchange. Extending this framework, \citet{UNIEJEWSKI2019} employed LASSO to compare univariate and multivariate specifications, finding that multivariate frameworks yielded no statistically significant forecasting edge, although they offered valuable insight into the underlying variable-selection patterns. Within a univariate setting, \citet{uniejewski2019weron} similarly applied LASSO to investigate variable selection underlying intraday dynamics in the German market. Finally, \citet{narajewski2020econometric}  analysed the ID$_3$-Price in the German Intraday Continuous electricity market using LASSO and Elastic Net and performed an out-of-sample, forecasting study. A well-documented limitation of these methods, however, is that they rely on the assumption that covariates are weakly correlated cross-sectionally and serially independent. When this assumption is violated, as is frequently the case in economic and financial datasets, and in particular in EPF, the ability of penalised-regression methods to consistently recover the true model is undermined, and they tend instead to retain a large number of variables with only marginal individual contribution to forecast accuracy.
	
	This limitation has motivated two distinct strands of the high-dimensional literature aimed at improving variable selection under strong cross-sectional dependence. The first strand, led by \citet{FarmSelect}, addresses the problem directly by decorrelating covariates before selection takes place. The proposed strategy, Factor-Adjusted Regularized Model Selection (FarmSelect), proceeds in two steps. First, it learns the parameters of an approximate factor model and identifies the highly correlated, low-rank component driving the covariates. This transforms the selection problem from one involving highly correlated covariates into one involving their weakly correlated idiosyncratic components together with the estimated factors. Second, FarmSelect solves a regularized profile likelihood problem using these transformed covariates as predictors. While this approach mitigates the effects of covariate correlation, it continues to rely on regularization, and therefore inherits the non-negligible false positive rates that such methods exhibit in the presence of pseudo-signals, that is, covariates which do not belong to the data-generating process (DGP) but are correlated with signal covariates that do. Such pseudo-signal effects arise frequently in macroeconomic and financial time series applications, many of which are driven in part by a small number of latent common factors. 
	
	The second strand comprises multiple testing approaches, which rely on marginal or conditional test statistics rather than shrinkage penalties to select variables. 
     This literature is led by the one covariate at a time multiple testing (OCMT) approach of \citet{OCMT}, later generalized by \citet{gocmt} to accommodate settings in which all covariates are correlated with the signals. Both approaches assess the statistical significance of each regressor through multiple testing and derive sharp bounds for the associated test statistics under the null hypothesis of no effect.
    At each stage, all regressors whose test statistics exceed the corresponding threshold are retained, yielding a decision rule that acts as a sharp filter for pure noise covariates while retaining those with genuine explanatory power. Although these methods offer a viable alternative to regularization, they too are known to suffer from non-negligible false positive rates in the presence of pseudo-signals.
	
	More recently, \citet{BMT} and \citet{BMTGLM} proposed Boosting with Multiple Testing (BMT), a variable selection procedure that combines stepwise forward variable addition with a family-wise multiple testing stopping rule. The key distinguishing feature of BMT is that, at each stage, only a single regressor is admitted to the model, even when several candidates appear individually significant. Rather than retaining all regressors found significant in marginal tests, as in OCMT, the procedure selects the most significant regressor conditional on those already included, updating the specification one variable at a time. At each subsequent stage, the remaining covariates are re-tested conditional on the expanded model, and again at most one additional regressor is admitted. This conservative, one-at-a-time admission rule has been found to deliver highly parsimonious specifications without sacrificing predictive accuracy, a property that motivates its use, alongside the methods discussed above, in the present paper.
	
	In this paper, we conduct an empirical evaluation comparing the out-of-sample forecasting performance of two traditional regularization techniques, specifically LASSO and Elastic Net, against four alternative screening frameworks, namely FarmSelect, OCMT, GOCMT, and BMT.  Our primary interest lies in introducing multiple testing approaches to the electricity price forecasting (EPF) context, where, to the best of our knowledge, they have not previously been employed; FarmSelect, a factor-adjusted regularization method, is included alongside them as a further screening-based benchmark. Our analysis draws on extensive spot price datasets from six major European electricity markets: Germany-Luxembourg, France, Spain, Norway (Oslo area), Finland, and Sweden (Stockholm area). Germany–Luxembourg, France, and Spain are among Europe’s largest electricity markets, while the inclusion of the Nordic areas extends the analysis to markets in which prices are driven by a different balance of fundamentals than those in continental Europe. In particular, Nordic electricity prices are strongly influenced by hydrological conditions due to the prominent role of hydropower in the energy mix, especially in hydro-dominated Norway. This combination allows us to assess the robustness of the methods across markets driven by different fundamentals. Predictive accuracy is formally assessed using the equal-accuracy test of \citet{DieboldMariano}. 
    
    We find that, while LASSO and elastic net exhibit highly similar forecasting performance and outperform FarmSelect, OCMT, and GOCMT on average, BMT achieves statistically comparable accuracy to these shrinkage benchmarks. Crucially, BMT delivers this competitive predictive power while maintaining a highly parsimonious specification, using less than one-tenth of the variables selected by the shrinkage benchmarks.  BMT therefore offers a viable alternative for optimizing variable selection and ensuring model simplicity without sacrificing predictive accuracy, all while incurring substantially lower computational cost than the regularization methods considered.
	
	The remainder of the paper is organized as follows. Section~\ref{sec:Data} introduces the spot price dataset. Section~\ref{sec: variable selection problem} describes the variable selection methods employed. Section~\ref{sec:Empirical} presents the results of the comparison across methods. Finally, Section~\ref{sec:conclusion} concludes with a brief discussion. 

\section{Data Description}\label{sec:Data}
Unlike other commodities, the electricity spot market is not continuously traded. Instead, it is a day-ahead market in which market participants submit bids and offers for the delivery of electricity for each delivery period of the following day. These bids and offers enter a blind auction known as the day-ahead auction in which supply and demand curves are constructed for each delivery period from the aggregated bids and offers. The electricity spot prices, or day-ahead prices, are then determined by matching supply and demand. In line with the majority of literature on European electricity markets, we use the terms spot price and day-ahead price interchangeably \citep{Uniejewski2016, Uniejewski2018, ZielWeron2018}.


The dataset in this study comprises hourly spot prices from six European price areas: Germany-Luxembourg (DE-LU), France (FR), Spain (ES), Norway (Oslo area, NO1), Finland (FI), and Sweden (Stockholm area, SE3). It also includes day-ahead forecasts of load (DA load) and generation from renewable energy sources (DA RES), where the latter is defined as the sum of day-ahead forecasts of onshore wind, offshore wind, and solar generation. Note that, since October 2025, the day-ahead market has comprised 96 15-minute delivery periods per day, replacing the previous 24 hourly periods. Consequently, to preserve the same frequency across the entire dataset, we downsample the prices observed since October 2025 by taking the mean across each hour, resulting in the dataset comprising hourly observations from 2020-01-01 to 2026-01-01. Following \citet{ZielWeron2018} and \citet{lago2021}, we transform days affected by daylight-savings time to contain 24 observations. On the transition to summer time, the missing hourly observation is imputed by linear interpolation between the two adjacent hours, while on the transition to standard time, the two observations corresponding to the repeated hour are replaced by their arithmetic mean. This procedure is applied to the spot price, DA load, and DA RES series. We additionally include fuel and carbon prices in the dataset, as these are key determinants of the marginal generation cost of conventional power plants. Specifically, we use front-month futures prices for the main European benchmarks for natural gas, coal, and oil, Dutch TTF Natural Gas, API2, and Brent Crude oil, respectively, traded on the Intercontinental Exchange (ICE). Moreover, we include the price of the December futures contracts on European Union emission allowances, EUA Dec, traded on the European Energy Exchange (EEX). We source the spot prices, DA load, and DA RES series from the ENTSO-E Transparency Platform \citep{entsoe_transparency}, and the fuel and carbon prices from Investing.com \citep{investing_com}, both of which are open-access. The API2 and Brent oil prices are quoted in USD, which we convert to EUR using the historical EUR/USD foreign exchange reference rate published by the European Central Bank \citep{ECBExchangeRates}.

We set the initial in-sample training window to 2020-01-01 to 2021-12-31, covering a total of 731 days. For the out-of-sample testing period, we consider the four years from 2022-01-01 to 2025-12-31, or a total of 1461 days. It is worth noting that the out-of-sample period includes 2022, a year marked by the European energy crisis and exceptionally high electricity prices and volatility. We follow \citet{Bara2023, Bara2024} in considering 2022 as our out-of-sample period. This makes the out-of-sample forecasting exercise more challenging and interesting. 

As is standard in the EPF literature and to mimic operational practice in, e.g., an energy trading company, we use a 731-day rolling window, corresponding to the length of the initial in-sample training window in the empirical application \citep{WERON20141030, ZielWeron2018}. Thus, each model is first calibrated using the initial in-sample training window, after which forecasts for all 24 hours of the following day are made. The window is then rolled forward one day, models are recalibrated, and forecasts for the next day are made. This is repeated until predictions for all 1461 days in the out-of-sample period have been made.

\begin{figure}[th!]
    \centering
    \includegraphics[width=\textwidth]{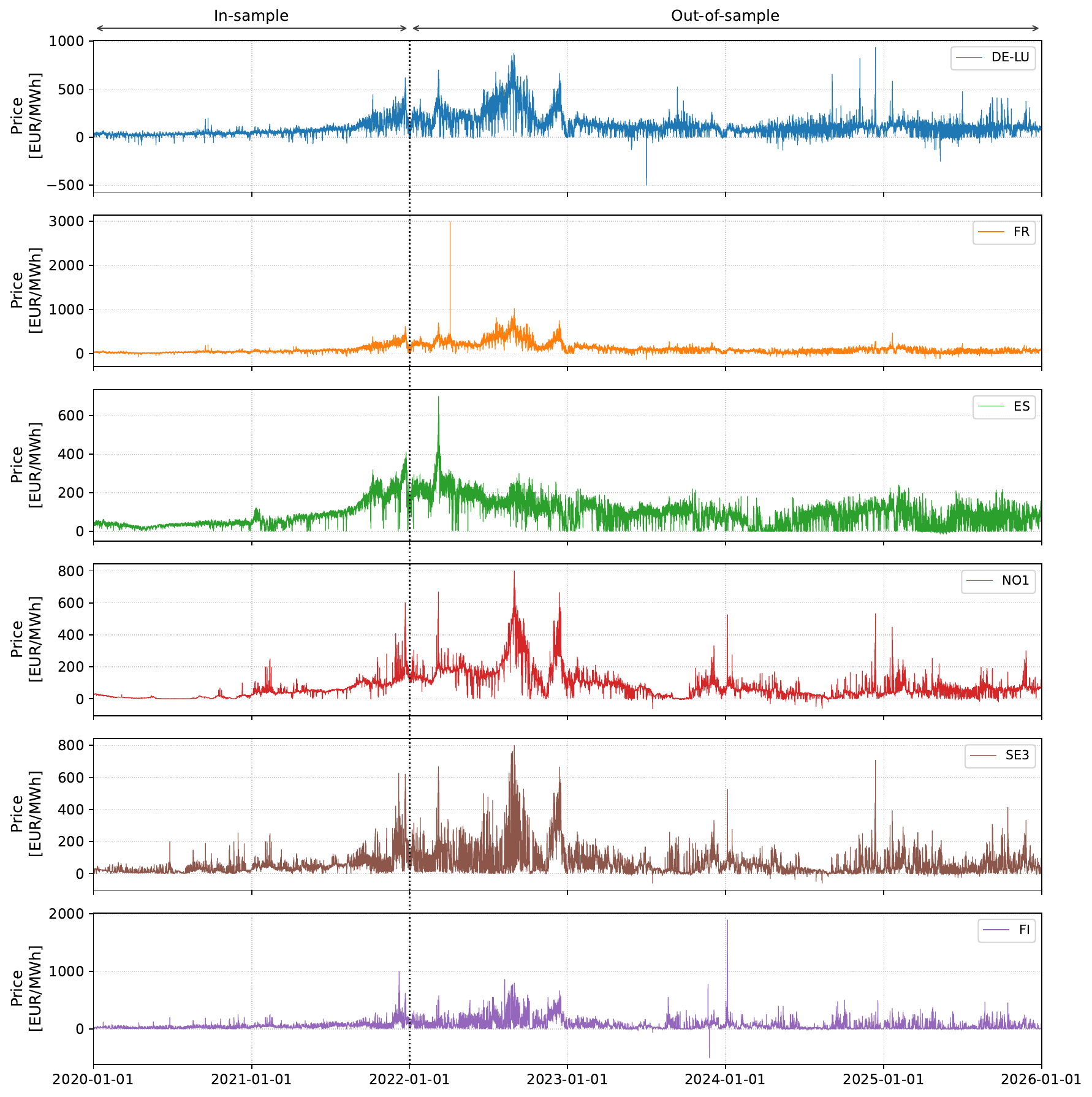}
    \caption{Day-ahead prices for the six considered price areas.}
    \label{fig: spot prices}
\end{figure}

Spot prices for the six areas are displayed in Figure~\ref{fig: spot prices}, from which it is clear that spot prices, regardless of the market we are considering, exhibit large positive and negative spikes. Other characteristics include daily, weekly, and annual seasonality. DA load display similar seasonal patterns, although without the extreme spikes observed in the spot prices, while DA RES is primarily characterized by daily and annual seasonality. In contrast, fuel and carbon prices behave more like random-walk-type processes with no seasonal structure.

DA load, DA RES, and fuel and carbon prices are displayed in Figure~\ref{fig: DA load}, Figure~\ref{fig: DA RES}, and Figure~\ref{fig: fuels} in \ref{appendix: data figures}, respectively. Notably, fuel and spot prices increased sharply across all price areas in 2022, as a result of the European energy crisis, which placed substantial pressure on electricity generation costs, driven, in part, by the significant reduction in natural gas supplies to Europe \citep{Ruhnau2023, SAETHER2024}.

\subsection{Variance stabilizing transformation}
\begin{figure}[ht!]
    \centering
    \includegraphics[width=\textwidth]{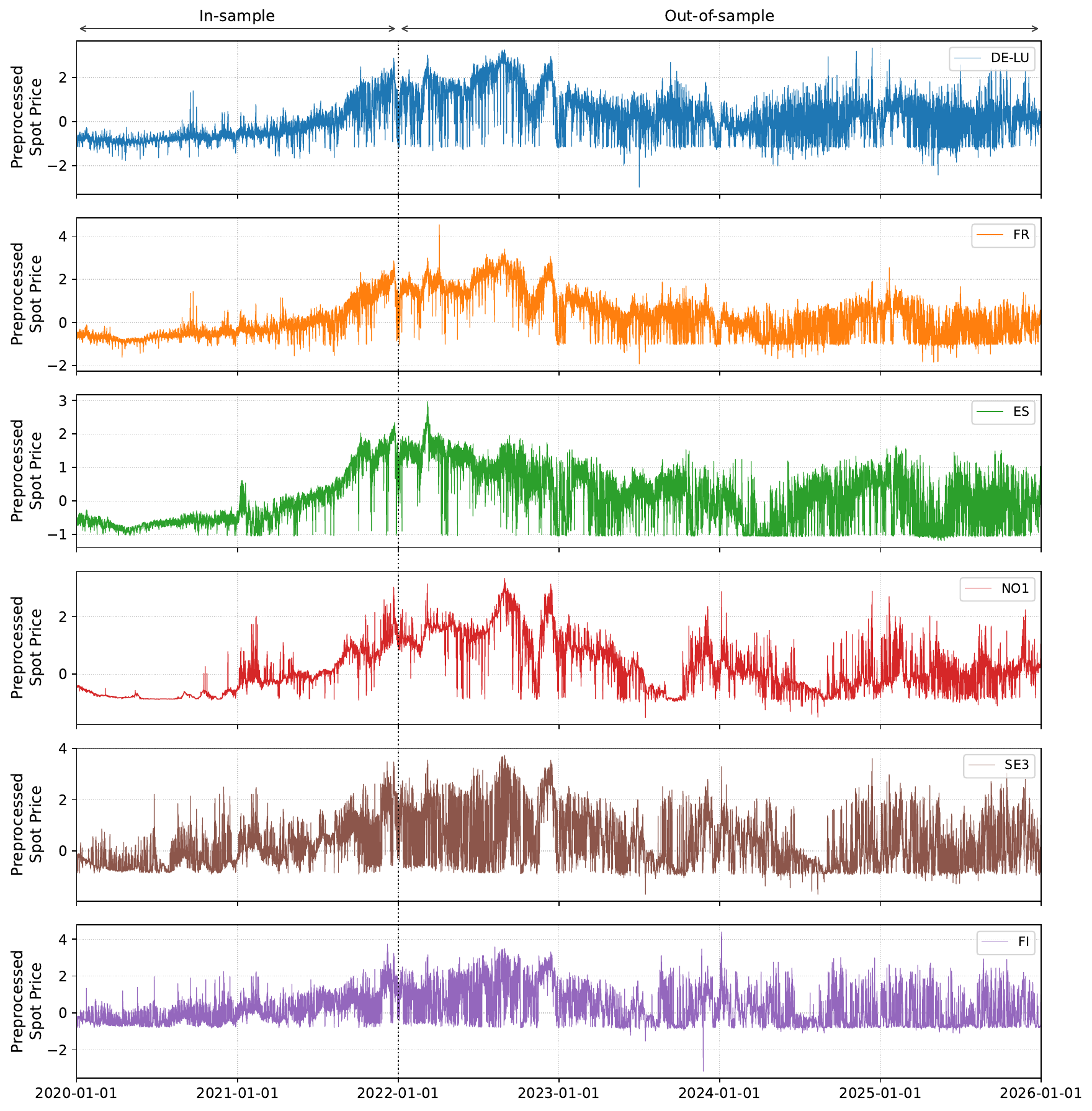}
    \caption{Preprocessed day-ahead prices for the six considered price areas using the area hyperbolic sine transformation.}
    \label{fig: spot preprocessed}
\end{figure}

Due to the spikes exhibited in the day-ahead prices, we follow common practices and recommendations in the EPF literature to enhance model performance by preprocessing the data using the area hyperbolic sine transformation:
$$
y_{t,h} = \operatorname{asinh}\left(\frac{p_{t,h}-a}{b}\right),
$$ 
where
$$
\operatorname{asinh}(x) \coloneqq \log(x+\sqrt{x^2+1}), \,\, x\in\R,
$$
and $ \frac{p_{t,h}-a}{b}$ represents standardized day-ahead prices for some shift and scale parameters $a$ and $b$. Following \citet{Uniejewski2018} and \citet{ZielWeron2018}, we set $a$ equal to the sample median and $b$ equal to the median absolute deviation (MAD) with the usual consistency correction. Preprocessed spot prices are displayed in Figure \ref{fig: spot preprocessed}, where we see that the area hyperbolic sine transformation has almost eliminated all price spikes, and dampened the impact of the European energy crisis.


Naturally, we transform forecasts $\hat{y}_{t,h}$ by the inverse transform, the hyperbolic sine function, to recover the day-ahead price forecasts
$$
\hat{p}_{t,h} = b\sinh{(\hat{y}_{t,h})} + a,
$$
which we use in the evaluation of the forecasts.

\section{Methodology}\label{sec: variable selection problem} 
Consider the high-dimensional linear regression model
\begin{equation}\label{eq: dgp general}
      y_{t} = \beta_0 + \sum_{i=1}^d\beta_i x_{t,i} + \varepsilon_t, \quad \text{for } t=1,\ldots, T,d\in\N,
\end{equation}
where  $(x_{t,i})_{i=1,\ldots,d}$ represent the (observed) covariates and $ \varepsilon_t$ is an error term. The main goal of the variable selection problem is to identify the  $k<d$,  unknown covariates in the high-dimensional regression model for which $\beta_i\neq0$. These are referred to as \emph{signal variables}, or simply \emph{signals}. Without loss of generality, we assume throughout this work that they are the first $k$ covariates in \eqref{eq: dgp general}. The remaining $d-k$ covariates are called \emph{inactive variables}. 
In particular, in the presence of $d$ candidate covariates we may write the DGP as
\begin{equation}\label{eq: dgp all variables}
    y_t = \beta_0 + \sum_{i=1}^d \mathbbm{1}(\beta_i \neq 0)\beta_i x_{t,i} + \varepsilon_t, \quad \text{for } t=1,\ldots, T,
\end{equation}
in which $\mathbbm{1}(A)$ denotes the usual indicator function.

As noted in the introduction, this study evaluates the performance of multiple testing approaches against penalized-regression methods. Since our interest lies in introducing these multiple testing frameworks to the electricity price forecasting context, we describe them in detail below, while technical details for the penalized-regression methods are deferred to \ref{appx_penalised_reggressions}. A key theoretical appeal of several of these approaches is that, under suitable conditions, the post-selection ordinary least squares (OLS) estimator based on the selected variables is asymptotically equivalent to the oracle estimator obtained under a known DGP.

\subsection{One Covariate at a Time Multiple Testing}
The One Covariate at a Time Multiple Testing (OCMT) procedure proposed by \citet{OCMT} formulates variable selection as an iterative multiple testing problem. At each stage, candidate covariates are tested individually using a multiplicity-adjusted critical value, conditional on the variables selected in previous stages. Once no additional variables are selected, the final model is estimated by OLS. 

To the best of our knowledge, the OCMT procedure is yet to be applied in the context of EPF, but \citet{OCMT} present a comprehensive simulation study 
directly comparing OCMT against LASSO, drawing promising conclusions that motivate its inclusion here. Although their empirical results indicate that neither method uniformly dominates the other across all settings, they find that OCMT frequently yields lower false-selection rates alongside competitive or lower forecast errors than LASSO. Moreover, its low computational cost is attractive in a forecasting setting in which variable selection must be repeated across forecast origins and delivery periods. We outline the procedure below using the setting and notation introduced above.

\paragraph{Stage 1} In the first stage, we consider $d$ regressions\footnote{We emphasize the fact that in the OCMT one may condition in the first stage on a set of known covariates. That is, instead of considering bivariate regressions, one may consider regressions of $y_t$ on a vector $z_t$ of known pre-selected variables, an intercept, and, one at a time, $x_{t,i}$ for $i = 1, 2,\ldots, d$: $$
y_t = z_t^\top a +  c_{i,(1)} + \phi_{i,(1)}x_{t,i} + e_{t,i,(1)}, \text{ for } t = 1, 2, \ldots, T.
$$
The procedure remains unchanged whether conditioning on a set of pre-selected variables or not.} of $y_t$ on an intercept and, one at a time, $x_{t,i}$ for $i = 1, 2,\ldots, d$, that is,
\begin{equation}\label{reg_xi_oneathtetime}
 y_t = c_{i,(1)} + \phi_{i,(1)}x_{t,i} + e_{t,i,(1)}, \text{ for } t = 1, 2, \ldots, T,   
\end{equation}
and denote the $\mathbf{t}$-statistic of $\phi_{i,(1)}$ by $\mathbf{t}_{\hat\phi_{i,(1)}}$ where $\hat\phi_{i,(1)}$ represents the OLS estimator of $\phi_{i,(1)}$. In this stage, the $x_i$ covariate is selected whenever 
$$
\lvert{\mathbf{t}_{\hat\phi_{i,(1)}}\rvert} > c_p(d,\delta_1).
$$
Here $c_p(d,\delta_1)$ denotes a critical value function defined as
\begin{equation}\label{eq: critical value function}
    c_p(d,\delta_1) := \Phi^{-1}\left( 1 - \frac{p}{2f(d,\delta_1)} \right),
\end{equation}
where $\Phi^{-1}$ is the inverse of the c.d.f. of standard normal distribution, $f(d,\delta) = cd^{\delta_1}$ for $c>0$, $\delta_1\geq1$, and $0<p<1$ is the significance level of the individual tests. Let $\mathcal{I}^s_{(1)}\subseteq\{1, \ldots, d\}$ be the index set of the selected variables and define $X_{(1)}^s:=(x_i)_{i\in\mathcal{I}^s_{(1)}}$. In addition, set $X_{(1)}:=X_{(1)}^s$, $\mathcal{N}_{(1)} := \{1,2,\ldots,d\} \setminus \mathcal{I}_{(1)}$, and $k_{(1)} := k_{(1)}^s$ where $k_{(1)}^s$ stands for the number of selected variables in this stage. 

\paragraph{Stage $j>1$} In stage $j = 2,3,\ldots$ we consider $n- k_{(j-1)}$ regressions of $y_t$ on an intercept and $x_{t,i}$, one at a time, for $i \in \mathcal{N}_{(j-1)}$, conditional on the variables in $X_{(j-1)}$, i.e.
$$
y_t = \beta_{(j)}\cdot X_{(j-1),t} +c_{i,(j)}+ \phi_{i,(j)}x_{t,i} + e_{t,i,(j)}, \text{ for } t=1, 2, \ldots, T.
$$
As in the previous stage, the covariate $x_i$ with $i\in\mathcal{N}_{(j-1)}$ is added to the set of already selected variables from stage $j-1$ whenever 
$$
\lvert{\mathbf{t}_{\hat\phi_{i,(j)}}\rvert} > c_p(d,\delta_2), \,\,\,\delta_2>\delta_1,
$$
in which $\mathbf{t}_{\hat\phi_{i,(j)}}$ denotes once again the $\mathbf{t}$-statistic of $\phi_{i,(j)}$ with $\hat\phi_{ i,(j)}$ being the OLS estimator of $\phi_{i,(j)}$. Similarly, we denote by $k_{(j)}^s$ the number of selected variables in stage $j$, the index set of the selected variables by $\mathcal{I}^s_{(j)}$, and the observation matrix of the selected variables by $X_{(j)}^s \in \R^{T \times k_{(j)}^s}$. We then update the total number of selected variables, the observation matrix of selected variables, the index set of selected variables, and the index set of variables not yet selected according to
\begin{alignat*}{2}
            X_{(j)} &= (X_{(j-1)}, X_{(j)}^s), \qquad k_{(j)} &&=  k_{(j-1)} + k_{(j)}^s\\
            \mathcal{I}_{(j)} &= \mathcal{I}_{(j-1)} \cup \mathcal{I}_{(j)}^s, \qquad\mathcal{N}_{(j)} &&= \{1,\ldots,n\} \setminus \mathcal{I}_{(j)}.
\end{alignat*}
If $k_{(j)}^s >0$ we move onto stage $j+1$, whereas if $k_{(j)}^s = 0$, i.e. no variables are selected in stage $j$, the procedure terminates and the total number of selected variables is set to $k_{(j-1)}$. In case we terminate the procedure, the last step consists of an OLS regression of $y$ on the selected variables in $X_{(j-1)}$.

\subsection{Generalized OCMT} 
A modified version of the OCMT procedure utilizing the factor model representation in \eqref{eq: factor model} to reduce correlation among covariates is the generalized OCMT (GOCMT) proposed by \citet{gocmt}. The proposed method builds upon the OCMT by conditioning on a set of pre-selected factors in every stage, thereby accounting for the common correlation across the covariates. Specifically, in the first stage, we consider $d$ regressions of the form
$$
y_t = c_{i,(1)} + \hat f_t\cdot \gamma_{i,(0)} + \phi_{i,(1)}x_{t,i} + e_{t,i,(1)}, \text{ for } t = 1, 2, \ldots, T,
$$
and the variables are selected according to the same multiple testing estimator as in OCMT. Similarly, for subsequent stages $j=2,3,\ldots$ we condition on the factors and the variables selected in previous stages: for $i\in\mathcal{N}_{(j-1)}$ estimate
$$
y_t= c_{i,(j)} + \hat f_t\cdot\gamma_{i,(j)}+\beta_{i,(j)}\cdot X_{(j-1),t}+\phi_{i,(j)}x_{t,i}+e_{t,i,(j)}, \text{ for } t = 1, 2, \ldots, T.
$$
Selection, updating, and termination then proceed as in the standard OCMT procedure.

\subsection{Boosting with Multiple Testing}
Boosting with Multiple Testing (BMT), proposed by \citet{BMT} and \citet{BMTGLM}, builds on OCMT by combining its multiple testing framework with forward stepwise variable selection. Whereas OCMT selects all candidate variables that pass the multiple testing filter, BMT selects only the covariate with the \textit{largest} absolute $\mathbf{t}$-statistic. The remaining candidate variables are then moved onto subsequent stages, where they are retested conditional on the set of selected variables. As in the OCMT, the procedure then continues until no covariates pass the multiple testing filter.

The stepwise selection of BMT is intended to reduce the selection of {\it pseudo-variables} or {\it pseudos}, i.e. covariates with $\beta_i=0$ that nevertheless appear predictive because they are correlated with true signals. Once a signal has entered the model, the conditional significance of its pseudos may disappear, allowing BMT to approximate the true DGP better. This property is particularly relevant for EPF, where variables like lagged prices, load and renewable forecasts, and fuel prices are often strongly correlated, leading to a potentially large amount of pseudo-variables. Further support for this is provided by the simulation study conducted in \citet{BMT}, where the authors conclude that BMT has better performance in terms of variable selection than OCMT and penalised regression methods - especially in cases of high multicollinearity. Since BMT relies on the same conditional regressions and multiple testing critical values as OCMT, we only describe the modifications that distinguish the two procedures.

\paragraph{Stage 1}
As in the first stage of OCMT, consider the $d$ regressions\footnote{Again, in the BMT one may condition on known pre-selected variables.} in \eqref{reg_xi_oneathtetime}. The variable with the largest $\mathbf{t}$-statistic among all candidate variables is then selected, i.e. the BMT selects the variable in the first stage according to
$$
i_{(1)} = \underset{i_{(1)}\in\mathcal{N}_{(0)}}{\arg\max} |\mathbf{t}_{\hat\phi_{i,(1)}}|,
$$ 
where  $\mathcal{N}_{(0)}=\{1,2,\ldots,d\}$. The selected and remaining index sets, along with the matrix of selected variables after the first stage, are then given as
$$
\mathcal{I}_{(1)}:=\{i_{(1)}\}, \qquad \mathcal{N}_{(1)} := \mathcal{N}_{(0)}\setminus\{i_{(1)}\}, \qquad X_{(1)} := (x_{i_{(1)}}).
$$

\paragraph{Stage $j>1$}
At stage \(j=2,3,\ldots\), each remaining candidate variable is tested separately conditional on the variables already selected. Thus, for each
$i\in\mathcal{N}_{(j-1)}$, we estimate
$$
y_t = \beta_{(j)}\cdot X_{(j-1),t} +c_{i,(j)}+ \phi_{i,(j)}x_{t,i} + e_{t,i,(j)}, \text{ for } t=1, 2, \ldots, T,
$$
and compute the corresponding $\mathbf{t}$-statistic
$\mathbf{t}_{\hat\phi_{i,(j)}}$. Let
$$
\mathcal{C}_{(j)} := \left\{i\in\mathcal{N}_{(j-1)} : |\mathbf{t}_{\hat\phi_{i,(j)}}| > c_p\left(n_{(j-1)},\delta_2\right) \right\},
$$
where $n_{(j-1)}$ is the number of candidate variables remaining from stage $j-1$ and $c_p$ as in \eqref{eq: critical value function}.
If $\mathcal{C}_{(j)}=\varnothing$, the procedure terminates and the final model is estimated by OLS using the variables contained in $X_{(j-1)}$. Otherwise, BMT selects only the most significant variable among those that pass the multiple testing filter:
$$
i_{(j)} = \underset{{i_{(j)}\in\mathcal{C}_{(j)}}}{\arg\max} |\mathbf{t}_{\hat\phi_{i,(j)}}|.
$$
The selected and remaining index sets and the matrix of selected variables are then updated according to
$$
\mathcal{I}_{(j)} = \mathcal{I}_{(j-1)}\cup\{i_{(j)}\}, \qquad \mathcal{N}_{(j)} = \mathcal{N}_{(j-1)}\setminus\{i_{(j)}\}, \qquad X_{(j)} = \left(X_{(j-1)},x_{i_{(j)}}\right).
$$
The procedure subsequently moves to stage $j+1$.

\section{Empirical Results}\label{sec:Empirical}

In this section, we consider the day-ahead electricity spot price forecasting problem. We first describe the forecasting model and the candidate variables. We then apply the variable selection methods introduced in Section~\ref{sec: variable selection problem} and compare the resulting models with LASSO, Elastic Net, and FarmSelect. The comparison is based on both forecasting accuracy and model parsimony. Details of the competing methods are provided in \ref{appx_penalised_reggressions} and references therein.

Following \citet{ZielWeron2018}, we adopt a day-ahead forecasting framework in which the 24-dimensional daily price vector is modeled using 24 separate univariate equations, one for each hour $h=1,2,\ldots,24$. As emphasized by \citet{UNIEJEWSKI2019}, a key advantage of automated variable selection is the ability to start out with a large set of candidate variables, and let the selection procedure determine which variables should be retained in the final model. Accordingly, we use the following set of candidate variables for every hour and for each variable selection method:
\begin{itemize}
    \item Transformed spot prices observed the previous seven days: $y_{t-1},y_{t-2}, \ldots, y_{t-7}$, in which $y_t = (y_{t,1}, y_{t,2}, \ldots, y_{t,24})^\top$.
    \item DA load for day $t$, and the seven previous days: $x^{(L)}_t, x^{(L)}_{t-1}, \ldots, \allowbreak x^{(L)}_{t-7}$, where $x^{(L)}_t = (x^{(L)}_{t,1}, x^{(L)}_{t,2}, \ldots, \allowbreak x^{(L)}_{t,24})^\top$.
    \item DA RES for day $t$, and the seven previous days: $x^{(RES)}_t, x^{(RES)}_{t-1}, \ldots, x^{(RES)}_{t-7}$, where $x^{(RES)}_t = (x^{(RES)}_{t,1}, x^{(RES)}_{t,2}, \ldots, x^{(RES)}_{t,24})^\top$.
    \item The most recent closing price of the EUA Dec futures contract observed on day $t$: $x^{(EUA)}_{t-2}$.
    \item The most recent closing prices of the front-month TTF Gas, API2 Coal, and Brent oil futures contracts observed on day $t$: $x^{(Gas)}_{t-2}, x^{(Coal)}_{t-2}, x^{(Oil)}_{t-2}$.
    \item Weekday dummy encoding the weekdays from Tuesday through Sunday by setting all elements equal to zero except the element that identifies the day of the week: $z_t = (z_{t,1}, z_{t,2}, \ldots, z_{t,6})^\top$ where $z_t = (1, 0, 0, 0, 0, 0)^\top$ represents Tuesday, $z_t= (0, 1, 0, 0, 0, 0)^\top$ represents Wednesday, etc. 
\end{itemize}
In total, we employ $d=562$ candidate variables. For each delivery hour $h=1,\ldots,24$, we consider the high-dimensional linear forecasting specification for the transformed spot price
\begin{equation}\label{eq: our variable selection problem}
    \begin{aligned}
        y_{t,h} = &\beta_{h,0} + \sum_{j=1}^7\sum_{i=1}^{24}\beta_{h,24(j-1) + i}y_{t-j,i} + \sum_{j=0}^7\sum_{i=1}^{24}\beta_{h,168 + 24j+i}x^{(L)}_{t-j,i} \\
        &+\sum_{j=0}^7\sum_{i=1}^{24}\beta_{h,360 + 24j+i}x^{(RES)}_{t-j,i} + \sum_{i=1}^6 \beta_{h,552+i}z_{t,i} \\ &+ \beta_{h,559}x^{(Gas)}_{t-2} + \beta_{h,560}x^{(Coal)}_{t-2} + \beta_{h,561}x^{(Oil)}_{t-2}+ \beta_{h,562}x^{(EUA)}_{t-2} + \varepsilon_{t,h},
    \end{aligned}
\end{equation}
for $t=1,2,\ldots, T$. This forecasting model is estimated using each of the variable selection methods considered in this paper. Throughout the empirical study, we use the following implementation settings. For LASSO and FarmSelect, we follow \citet{lago2021} and use the least angle regression (LARS) of \citet{efron} to select the regularization parameter according to the Akaike Information Criterion (AIC). In Elastic Net, we set $M=50$ in the two-dimensional grid search described in \ref{appx_elasticnet} and select the regularization parameters based on the AIC. To estimate the factors in FarmSelect and GOCMT, we use principal component analysis (PCA) and use the modified eigenvalue-ratio criterion described in \ref{appx_farmselect} with $C_D=0$ and $\zeta_{max}=20$ to select the number of factors. For OCMT and GOCMT, we follow the empirical specification of \citet{OCMT} and set $\delta_1=1$ and $\delta_2=2$. For BMT, we use the same settings as in the empirical application of \citet{BMT} and set $\delta_1=\delta_2=1$. For all three multiple testing-based procedures, we use an individual-test significance level of $p=0.05$ and set $c=1$. To account for serial dependence, the test statistics at each stage of the multiple testing procedures are computed using heteroskedasticity- and autocorrelation-consistent (HAC) standard errors based on the Bartlett kernel and the automatic bandwidth selection rule of \citet{NeweyWest1994}. We evaluate the forecasting performance of each method in the out-of-sample test period using a series of commonly used evaluation metrics and statistical tests based on the equal-accuracy test of \citet{DieboldMariano} in the EPF literature.

\subsection{Evaluation metrics}
Following \citet{lago2021}, we use the standard forecast evaluation metrics, mean absolute error (MAE), and root mean squared error (RMSE):
\begin{align*}
        \operatorname{MAE} &= \frac{1}{24N}\sum_{t=1}^{N}\sum_{h=1}^{24} \left\vert \hat{p}_{t,h} - p_{t,h}  \right\vert \\
    \operatorname{RMSE} &= \sqrt{\frac{1}{24N}\sum_{t=1}^{N}\sum_{h=1}^{24} \left( \hat{p}_{t,h} - p_{t,h}  \right)^2},
\end{align*}
where $N$ denotes the number of days in the out-of-sample test period. The mean absolute percentage error (MAPE) is also identified by \citet{lago2021} as a commonly used evaluation metric, with its scale-free nature facilitating comparisons across price areas with different price levels. However, the MAPE has important limitations in the context of EPF. Firstly, regardless of the absolute forecast error, the MAPE explodes for prices near zero, and secondly, it is not defined for prices that are exactly zero - something that is common in the day-ahead market where prices tend to "stick" to zero.

To alleviate some of these problems, we consider the symmetric MAPE (sMAPE) defined in \citet{HYNDMAN2006} as
$$
\operatorname{sMAPE} = \frac{1}{24N}\sum_{t=1}^{N}\sum_{h=1}^{24} 2 \frac{\left\vert \hat{p}_{t,h} - p_{t,h}  \right\vert}{\left\vert \hat{p}_{t,h}  \right\vert + \left\vert p_{t,h}  \right\vert}.
$$

Lastly, following the suggestions of \citet{lago2021} for fair comparisons of the different methods, we also consider the relative MAE (rMAE), which normalizes the MAE by the MAE of a naive forecast:
$$
\operatorname{rMAE} = \frac{\displaystyle \frac{1}{24N}\sum_{t=1}^{N}\sum_{h=1}^{24} \left\vert \hat{p}_{t,h} - p_{t,h}  \right\vert}{\displaystyle \frac{1}{24N_d}\sum_{t=1}^{N}\sum_{h=1}^{24} \left\vert \hat{p}^{naive}_{t,h} - p_{t,h}  \right\vert}.
$$
To account for weekly seasonality in day-ahead prices, we choose
\begin{equation}
    \hat{p}^{naive}_{t,h} = p_{t-7,h},
\end{equation}
as the naive forecast.

Evaluation metrics for the different variable selection methods across the six price areas in the out-of-sample period are displayed in Table~\ref{tab: hourly metrics}. In addition to the above-listed metrics, we also report the average number of selected covariates of each method across both the 24 hours and days in the test period, which we denote by $\operatorname{\#vars}$.



Results in Table~\ref{tab: hourly metrics} show that BMT and Elastic Net provide the strongest overall forecast performance, with LASSO generally following closely. Indeed, the lowest value for every price-area–metric combination is obtained by either BMT or Elastic Net, with Elastic Net minimizing all four evaluation metrics in DE-LU and FR, although with BMT closely following in DE-LU. By contrast, BMT minimizes all four metrics in FI, NO1, and SE3. In ES, BMT obtains the lowest rMAE, MAE, and RMSE, while Elastic Net has a marginally lower sMAPE. Averaged equally across the six price areas, BMT and Elastic Net have the lowest rMAE at $0.501$, followed by LASSO at $0.514$.

\begin{figure}[t!]
    \centering
    \includegraphics[width=0.49\linewidth]{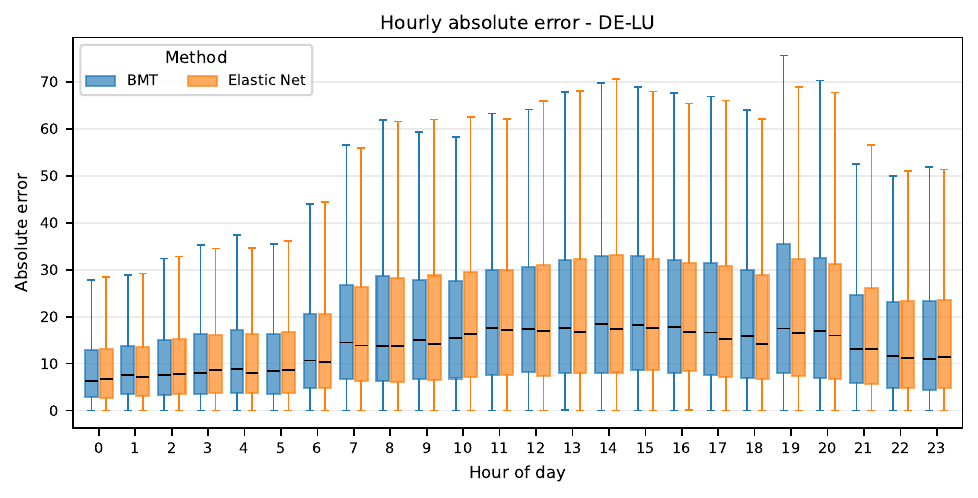}\includegraphics[width=0.49\linewidth]{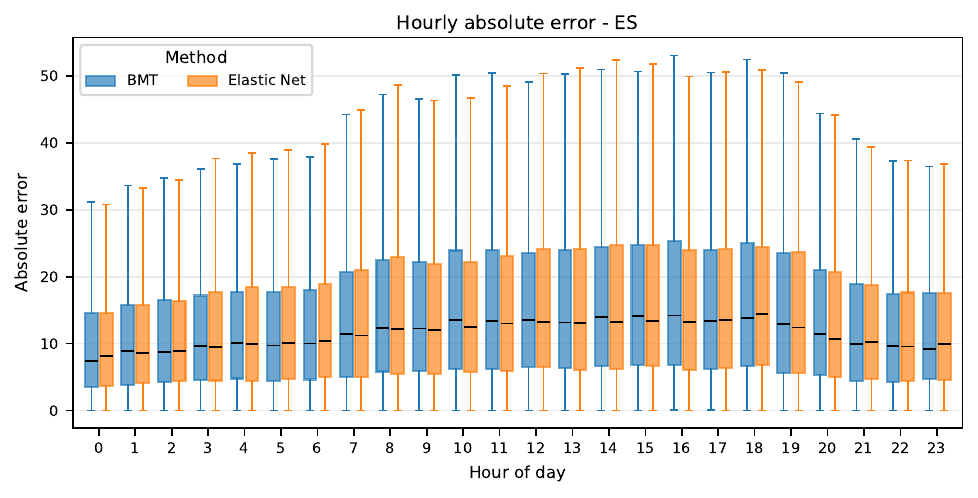}
    \includegraphics[width=0.49\linewidth]{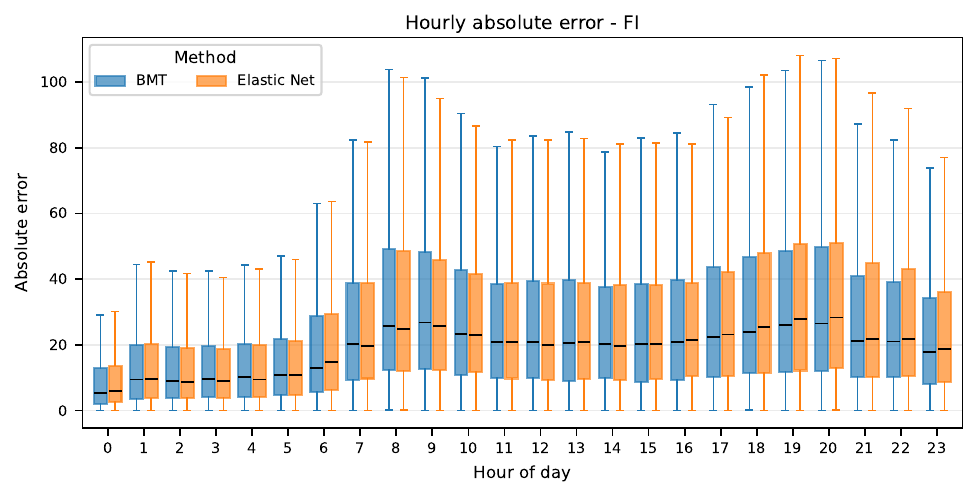}\includegraphics[width=0.49\linewidth]{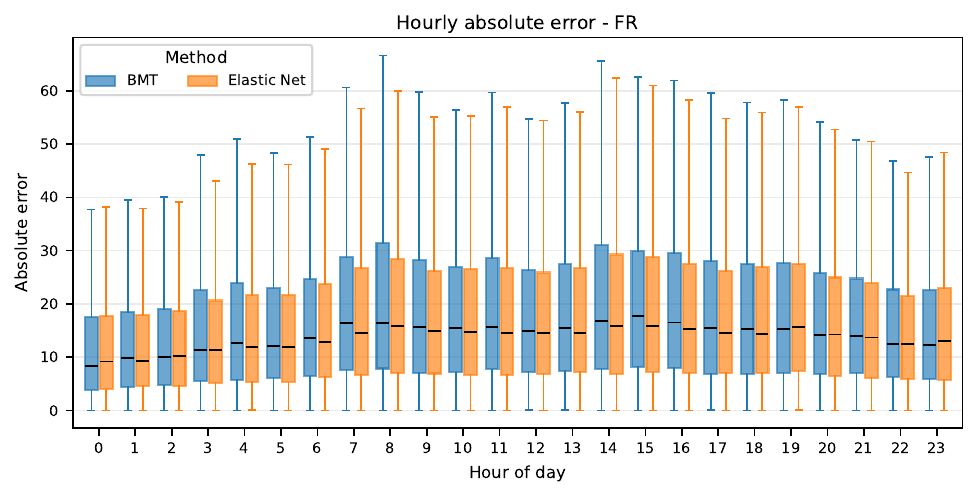}
    \includegraphics[width=0.49\linewidth]{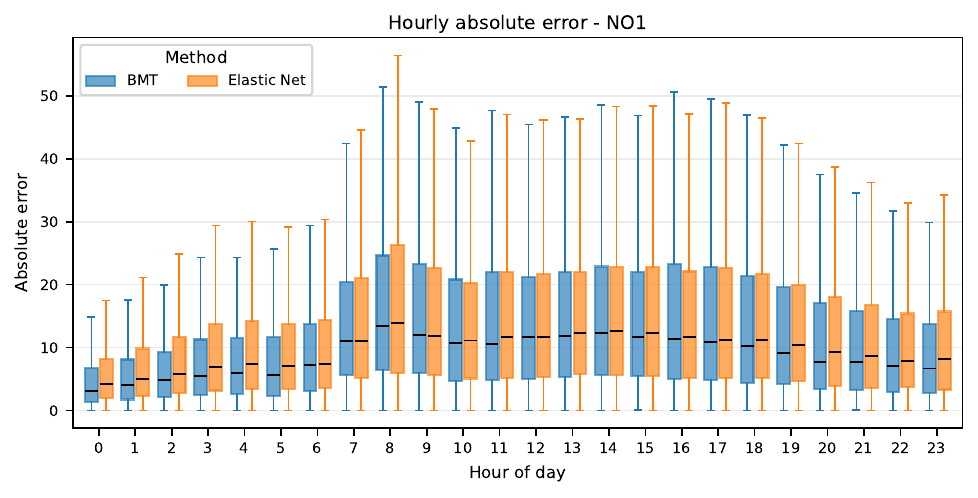}\includegraphics[width=0.49\linewidth]{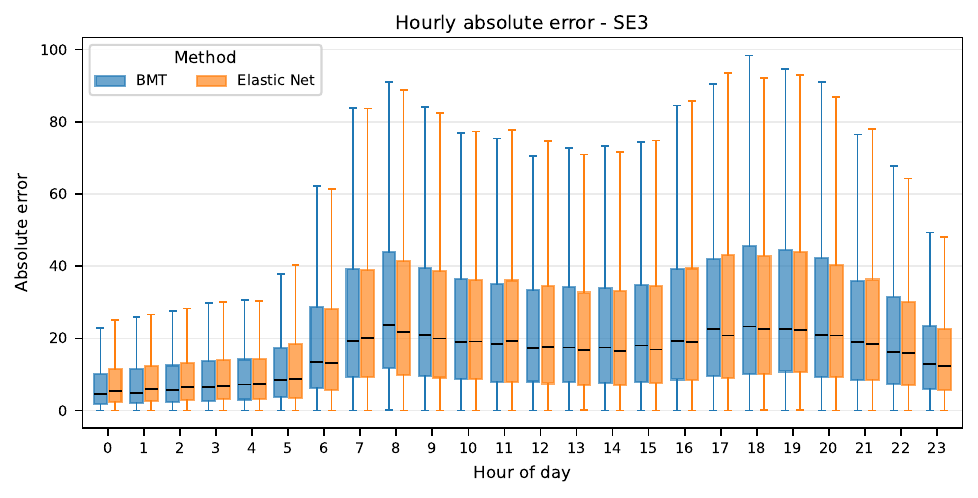}
    \caption{Boxplot of hourly absolute error for BMT and Elastic Net. }
    \label{fig: hourly AE}
\end{figure}

A particularly interesting result concerns the degree of sparsity of the different methods. BMT selects, on average, between approximately $4.5$ and $8.2$ variables across the six price areas. In comparison, Elastic Net selects between $88.2$ and $133$ variables, while LASSO selects between $107$ and $132$ variables. Depending on the price area and comparator, BMT therefore selects approximately $90\%$–$95\%$ fewer variables, or, equivalently, the specifications produced by the regularization-based methods are approximately 10 to 20 times larger. The combination of sparsity and forecast accuracy is especially pronounced in the Nordic price areas, where BMT produces the smallest specifications and the lowest evaluation metrics. The results for NO1 stand out in particular: BMT achieves the best performance on all four evaluation metrics while selecting fewer than five variables on average. By comparison, LASSO and Elastic Net select approximately $131$ and $105$ variables, respectively. Importantly, this substantial reduction in dimensionality does not come at the expense of forecast accuracy. Figure~\ref{fig: hourly AE} shows boxplots of the hourly absolute error for BMT and Elastic Net across the six price areas. At first glance, there does not appear to be a substantial difference between the two methods: both the median error and the interquartile range track each other closely at almost every hour of the day, in every market. The shared diurnal pattern, low errors overnight rising to a plateau during daytime and evening hours before tapering off, is essentially identical across methods, and the whiskers and occasional outliers reach comparable magnitudes for BMT and Elastic Net alike. This similarity holds regardless of overall price-level differences across markets, from the tighter error ranges observed in ES and NO1 to the wider ranges in FI and SE3.

The remaining methods generally provide less accurate forecasts. FarmSelect selects approximately as many variables as LASSO but yields higher evaluation metrics in every price area. OCMT and GOCMT generate the largest specifications, selecting approximately $257$–$305$ and $183$–$282$ variables, respectively, without a corresponding improvement in forecast accuracy; neither method attains the lowest value of any evaluation metric in any of the six markets. The nature of the variable selection problem in this study means that the candidate set of covariates contains several highly dependent variables, and as documented in \citet{BMT}, a downside of OCMT is its tendency to include several highly dependent variables in such a setting. To examine whether this occurs in the present application, we divide the candidate variables into five groups---Price lags, Load, RES, Fuels, and Weekdays---comprising lagged spot prices, DA load, DA RES, fuel and EUA prices, and weekday indicators, respectively. 

\begin{figure}[t!]
    \centering
    \includegraphics[width=\linewidth]{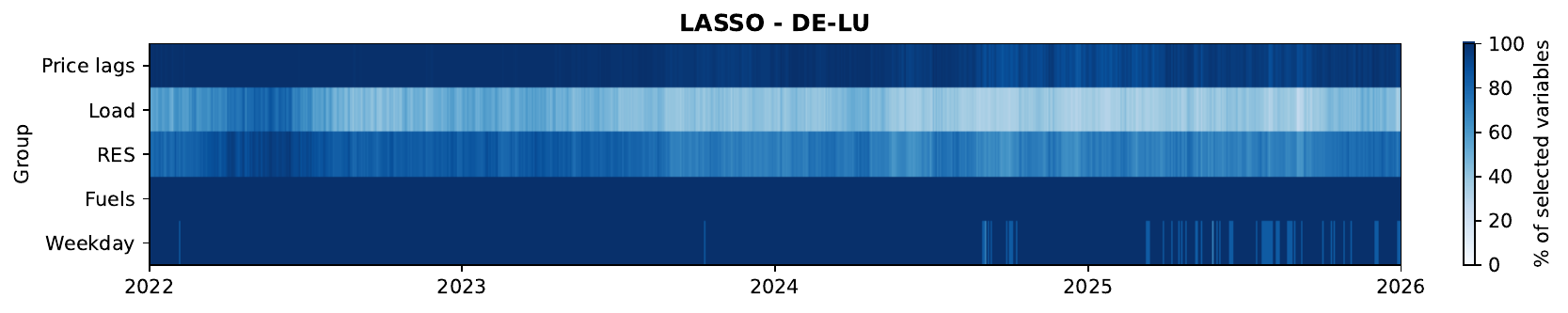}
    \includegraphics[width=\linewidth]{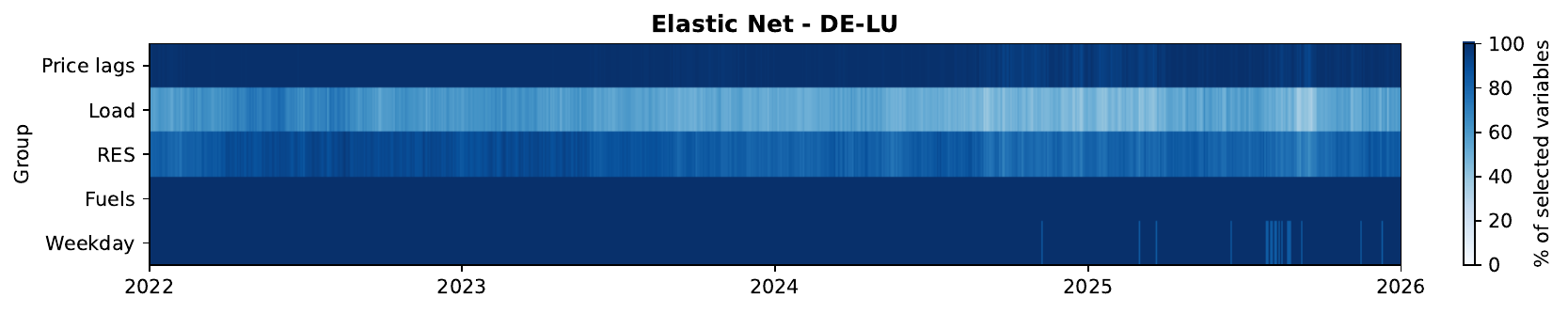}
    \includegraphics[width=\linewidth]{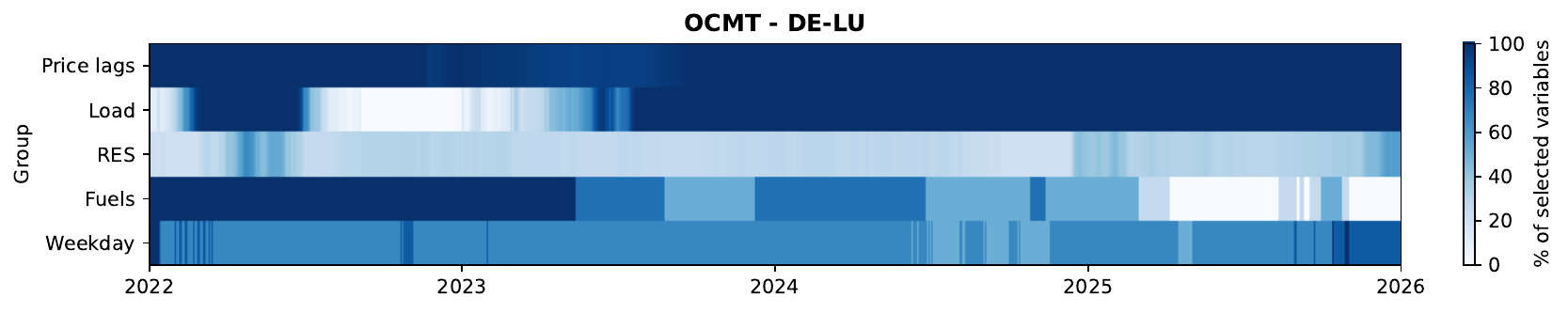}
    \includegraphics[width=\linewidth]{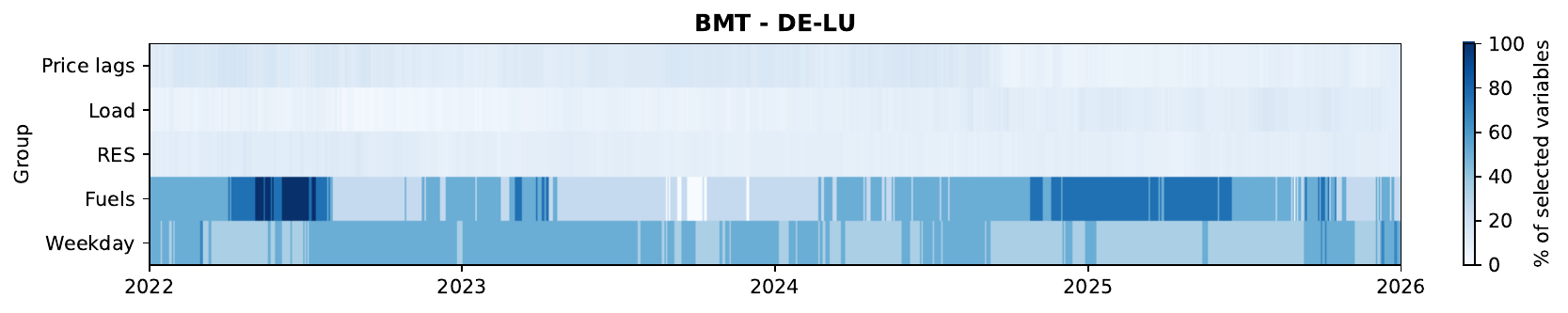}
    \includegraphics[width=\linewidth]{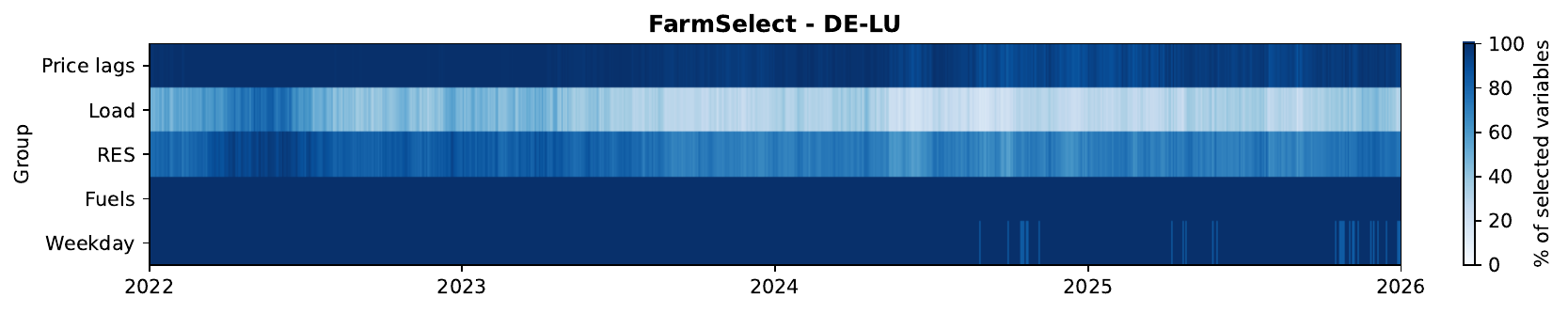}
    \includegraphics[width=\linewidth]{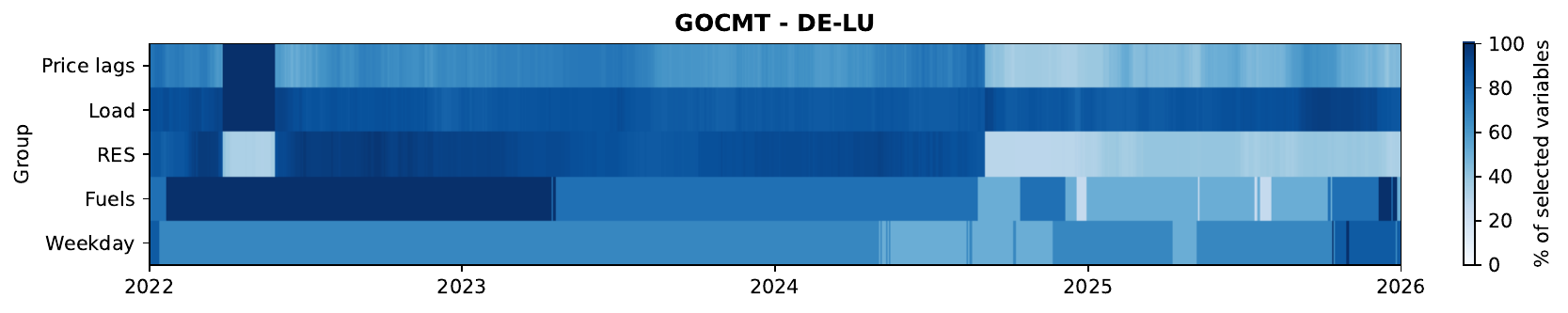}
    \caption{Heat map of the percentage of selected number of variables for each method from the groups Price lags, Load, RES, Fuels, and Weekday.}
    \label{fig: selected heatmaps}
\end{figure}

Figure~\ref{fig: selected heatmaps} displays the percentage of candidate variables selected from each group over the evaluation period for DE-LU. The heatmaps show that the large specifications produced by OCMT and GOCMT appear to be driven, at least in part, by their tendency to select clusters of highly dependent variables, consistent with the findings of \citet{BMT}. LASSO, Elastic Net, and FarmSelect also retain sizeable proportions of variables from these groups, consistent with the relatively large specifications reported in Table~\ref{tab: hourly metrics}. In contrast, the BMT only selects a small proportion of the variables from Price lags, Load, and RES groups, indicating that BMT's parsimonious specifications arise partly from avoiding the redundant selection of clusters of highly dependent variables. A similar picture is observed for the other five price areas, which are displayed in \ref{appendix: result figures} to conserve space.

\begin{table}
\centering
\caption{Forecast performance metrics by price area. Bold values indicate the lowest value within each price area. An rMAE below one indicates an improvement relative to the naive forecast.}
\label{tab: hourly metrics}
\begin{tabular}
{ll|ccccccc}\hline\hline
 &  & LASSO & Elastic Net & OCMT & BMT & FarmSelect & GOCMT \\
\hline
\multirow{5}{*}{\textbf{DE-LU}}
& rMAE & 0.443 & \textbf{0.426} & 0.579 & 0.428 & 0.678 & 0.543 \\
 & MAE & 20.633 & \textbf{19.837} & 26.971 & 19.953 & 31.577 & 25.312 \\
 & sMAPE [\%] & 30.011 & \textbf{29.683} & 35.106 & 29.946 & 42.862 & 34.659 \\
 & RMSE & 33.346 & \textbf{31.664} & 44.308 & 31.742 & 48.277 & 41.276 \\
 & \#vars & 107.257 & 98.942 & 280.415 & \textbf{8.107} & 105.167 & 248.298 \\
\midrule
\multirow{5}{*}{\textbf{ES}}
& rMAE & 0.520 & 0.497 & 0.677 & \textbf{0.496} & 0.654 & 0.613 \\
 & MAE & 16.282 & 15.543 & 21.189 & \textbf{15.521} & 20.467 & 19.169 \\
 & sMAPE [\%] & 37.443 & \textbf{36.866} & 44.095 & 36.971 & 41.022 & 41.715 \\
 & RMSE & 23.135 & 21.943 & 29.664 & \textbf{21.816} & 29.930 & 26.701 \\
 & \#vars & 131.794 & 104.266 & 285.662 & \textbf{6.893} & 129.063 & 216.891 \\
\midrule
\multirow{5}{*}{\textbf{FI}}
& rMAE & 0.535 & 0.537 & 0.693 & \textbf{0.528} & 0.631 & 0.654 \\
 & MAE & 30.041 & 30.191 & 38.940 & \textbf{29.682} & 35.470 & 36.757 \\
 & sMAPE [\%] & 68.577 & 69.042 & 78.893 & \textbf{68.411} & 83.102 & 77.432 \\
 & RMSE & 55.098 & 55.163 & 73.983 & \textbf{54.495} & 61.516 & 65.665 \\
 & \#vars & 115.073 & 133.525 & 257.604 & \textbf{7.662} & 111.853 & 182.545 \\
\midrule
\multirow{5}{*}{\textbf{FR}}
& rMAE & 0.491 & \textbf{0.470} & 0.691 & 0.497 & 0.806 & 0.664 \\
 & MAE & 19.694 & \textbf{18.840} & 27.715 & 19.936 & 32.320 & 26.600 \\
 & sMAPE [\%] & 34.955 & \textbf{34.401} & 44.294 & 35.342 & 48.875 & 43.187 \\
 & RMSE & 33.593 & \textbf{32.457} & 43.684 & 34.056 & 51.975 & 41.203 \\
 & \#vars & 121.862 & 88.135 & 304.591 & \textbf{8.228} & 118.721 & 282.045 \\
\midrule
\multirow{5}{*}{\textbf{NO1}}
& rMAE & 0.539 & 0.517 & 0.762 & \textbf{0.502} & 0.707 & 0.624 \\
 & MAE & 16.039 & 15.392 & 22.682 & \textbf{14.958} & 21.057 & 18.577 \\
 & sMAPE [\%] & 32.227 & 31.473 & 38.924 & \textbf{29.861} & 45.178 & 35.184 \\
 & RMSE & 27.868 & 26.743 & 50.617 & \textbf{26.667} & 33.054 & 34.718 \\
 & \#vars & 131.423 & 104.662 & 256.628 & \textbf{4.485} & 128.545 & 201.249 \\
\midrule
\multirow{5}{*}{\textbf{SE3}}
& rMAE & 0.555 & 0.560 & 0.745 & \textbf{0.554} & 0.656 & 0.709 \\
 & MAE & 25.137 & 25.399 & 33.754 & \textbf{25.091} & 29.715 & 32.141 \\
 & sMAPE [\%] & 59.184 & 59.713 & 73.173 & \textbf{58.769} & 74.485 & 70.075 \\
 & RMSE & 43.468 & 43.899 & 60.598 & \textbf{42.249} & 50.248 & 55.364 \\
 & \#vars & 107.384 & 129.928 & 291.508 & \textbf{6.876} & 102.976 & 227.834 \\
\hline \hline
\end{tabular}
\end{table}

\subsubsection{Evaluation metrics for daily baseload prices}
In addition to the hourly analysis, we also consider forecasts of the daily baseload price, obtained by averaging the 24 hourly price forecasts. The daily baseload price represents the price of a constant delivery of electricity throughout the day. Previous studies have shown that forecasting the hourly prices before aggregation leads to more accurate baseload forecasts than forecasting the daily aggregate directly \citep{Raviv2015,MaciejowskaWeron2016}.  Daily baseload price forecasts are particularly relevant for electricity trading. In price areas where Base Day Futures are traded, including DE-LU, FR, and ES, traders can compare the forecast of the daily baseload price with the quoted futures price for delivery on the following day. An expected daily baseload price above the futures price may motivate a long position, whereas an expected daily baseload price below the quoted market price may motivate a short position. Since these contracts are financially settled against an index derived from the average day-ahead prices of the following day \citep{EEX2026}, the accuracy of the aggregated forecast is directly related to traders' assessment of potential contract payoff and decision-making in the market.

We evaluate the performance of the different methods based on the same evaluation metrics as for the hourly forecasts, but adjusted to daily values. Let $p_t = \frac{1}{24}\sum_{h=1}^{24} p_{t,h}$ and $\hat{p}_t = \frac{1}{24}\sum_{h=1}^{24} \hat{p}_{t,h}$ denote the daily baseload price on day $t$ and the forecast of the daily baseload price on day $t$, respectively. Then we consider the following daily evaluation metrics:
\begingroup
\allowdisplaybreaks
\begin{align*}
    \operatorname{MAE_t} &= \frac{1}{N}\sum_{t=1}^{N} \left\vert \hat{p}_{t} - p_{t}  \right\vert, \\
    \operatorname{RMSE_t} &= \sqrt{\frac{1}{N}\sum_{t=1}^{N} \left( \hat{p}_{t} - p_{t}  \right)^2}, \\
    \operatorname{sMAPE_t} &= 2\frac{1}{N}\sum_{t=1}^{N} \frac{\left\vert \hat{p}_{t} - p_{t}  \right\vert}{\left\vert \hat{p}_{t}  \right\vert + \left\vert p_{t}  \right\vert},\\
    \operatorname{rMAE_t} &= \frac{\displaystyle \frac{1}{N}\sum_{t=1}^{N} \left\vert \hat{p}_{t} - p_{t}  \right\vert}{\displaystyle \frac{1}{N}\sum_{t=1}^{N} \left\vert \hat{p}^{naive}_{t} - p_{t}  \right\vert},
\end{align*}
\endgroup
where
$$
\hat{p}^{naive}_{t} = p_{t-7}.
$$
Evaluation metrics for forecast performance of daily baseload prices are reported in Table~\ref{tab: daily metrics}. Here, BMT obtains the lowest value for all four evaluation measures in all price areas except FR, where Elastic Net performs the best, followed by LASSO.

Overall, BMT ranks first in 20 of the 24 price-area–metric combinations, while Elastic Net ranks first in the remaining four. Compared with the hourly evaluations, BMT replaces Elastic Net as the best-performing method across all four metrics in DE-LU and also takes the remaining metric in ES. Averaged equally across the six price areas, BMT obtains the lowest $\operatorname{rMAE}_{t}$ at $0.408$, followed by Elastic Net at $0.417$ and LASSO at $0.424$.

The remaining methods continue to provide less accurate forecasts. FarmSelect is outperformed by every other method across all price areas. For the OCMT-type methods, the picture is the same as for the hourly evaluation metrics, with GOCMT outperforming OCMT across all price areas.

\begin{table}[t!]
    \centering
    \caption{Forecasting performance by price area for daily baseload prices. Bold values indicate the lowest value within each price area. An rMAE below one indicates an improvement relative to the naive forecast.}
    \label{tab: daily metrics}
\begin{tabular}{ll|cccccc}
\hline \hline
 &  & LASSO & Elastic Net & OCMT & BMT & FarmSelect & GOCMT \\
\midrule
\multirow{4}{*}{\textbf{DE-LU}}
& $\operatorname{rMAE_t}$ & 0.367 & 0.355 & 0.469 & \textbf{0.347} & 0.680 & 0.420 \\
 & $\operatorname{MAE_t}$ & 15.151 & 14.673 & 19.384 & \textbf{14.320} & 28.080 & 17.352 \\
 & $\operatorname{sMAPE_t}$ [\%] & 17.966 & 17.836 & 21.892 & \textbf{17.608} & 34.454 & 21.245 \\
 & $\operatorname{RMSE_t}$ & 25.315 & 24.502 & 31.968 & \textbf{23.861} & 42.827 & 28.115 \\
\midrule
\multirow{4}{*}{\textbf{ES}}
& $\operatorname{rMAE_t}$ & 0.422 & 0.403 & 0.548 & \textbf{0.391} & 0.586 & 0.467 \\
 & $\operatorname{MAE_t}$ & 11.744 & 11.218 & 15.256 & \textbf{10.892} & 16.326 & 13.017 \\
 & $\operatorname{sMAPE_t}$ [\%] & 21.943 & 21.274 & 27.371 & \textbf{21.028} & 25.717 & 24.838 \\
 & $\operatorname{RMSE_t}$ & 16.816 & 16.069 & 21.417 & \textbf{15.625} & 25.409 & 18.378 \\
\midrule
\multirow{4}{*}{\textbf{FI}}
& $\operatorname{rMAE_t}$ & 0.450 & 0.456 & 0.567 & \textbf{0.443} & 0.592 & 0.524 \\
 & $\operatorname{MAE_t}$ & 22.844 & 23.145 & 28.785 & \textbf{22.451} & 30.051 & 26.579 \\
 & $\operatorname{sMAPE_t}$ [\%] & 53.922 & 54.437 & 62.474 & \textbf{51.508} & 72.159 & 61.122 \\
 & $\operatorname{RMSE_t}$ & 40.416 & 40.769 & 50.400 & \textbf{40.354} & 48.722 & 44.983 \\
\midrule
\multirow{4}{*}{\textbf{FR}}
& $\operatorname{rMAE_t}$ & 0.387 & \textbf{0.375} & 0.538 & 0.388 & 0.807 & 0.510 \\
 & $\operatorname{MAE_t}$ & 13.900 & \textbf{13.473} & 19.351 & 13.961 & 29.006 & 18.345 \\
 & $\operatorname{sMAPE_t}$ [\%] & 21.027 & \textbf{20.837} & 29.466 & 20.967 & 39.547 & 28.442 \\
 & $\operatorname{RMSE_t}$ & 20.554 & \textbf{19.915} & 28.065 & 20.824 & 44.287 & 25.877 \\
\midrule
\multirow{4}{*}{\textbf{NO1}}
& $\operatorname{rMAE_t}$ & 0.446 & 0.437 & 0.639 & \textbf{0.413} & 0.668 & 0.504 \\
 & $\operatorname{MAE_t}$ & 12.092 & 11.856 & 17.328 & \textbf{11.197} & 18.103 & 13.657 \\
 & $\operatorname{sMAPE_t}$ [\%] & 23.636 & 23.512 & 30.145 & \textbf{21.972} & 39.402 & 26.095 \\
 & $\operatorname{RMSE_t}$ & 19.704 & 19.114 & 35.374 & \textbf{18.503} & 26.590 & 23.214 \\
\midrule
\multirow{4}{*}{\textbf{SE3}}
& $\operatorname{rMAE_t}$ & 0.470 & 0.474 & 0.610 & \textbf{0.466} & 0.613 & 0.574 \\
 & $\operatorname{MAE_t}$ & 19.167 & 19.321 & 24.901 & \textbf{19.024} & 24.997 & 23.401 \\
 & $\operatorname{sMAPE_t}$ [\%] & 44.026 & 44.309 & 56.090 & \textbf{42.289} & 63.819 & 54.884 \\
 & $\operatorname{RMSE_t}$ & 31.359 & 31.524 & 40.780 & \textbf{30.358} & 39.567 & 36.257 \\
\hline \hline
\end{tabular}
\end{table}

\subsection{Diebold-Mariano test}
In addition to the evaluation metrics, we also evaluate whether any differences in forecast accuracy between methods are statistically significant. For this purpose, we use the Diebold--Mariano (DM) test \citep{DieboldMariano} that is commonly employed in the EPF literature \citep{lago2021, ZielWeron2018}. Two versions of the DM test exist for EPF: a univariate version and a multivariate version. The univariate version performs 24 individual tests; one for each hour of the day. For fixed $h$, the univariate version uses the loss differential series defined by
$$
\Delta_{t,h}^{A,B}=\left|\varepsilon_{t,h}^{A}\right|^p - \left|\varepsilon_{t,h}^{B}\right|^p, \quad t = 1,2,\ldots,N,
$$
where $\varepsilon_{t,h}^{Z}=p_{t,h}-\hat{p}_{t,h}^{Z}$ denotes the forecast error of model $Z$, and $p=1, 2$. Introduced in \citet{ZielWeron2018}, the multivariate DM test is performed jointly across all 24 delivery periods. The multivariate DM test uses the multivariate loss differential series defined as
\begin{equation}\label{eq: multivariate loss differential}
    \Delta_t^{A,B} = \| \varepsilon_t^{A} \|_p - \| \varepsilon_t^{B} \|_p,
\end{equation}
where $\varepsilon_t^{Z} = (\varepsilon_{t,1}^{Z}, \ldots, \varepsilon_{t,24}^{Z})^\top$ is a 24-dimensional vector of prediction errors for model $Z$ on day $t$, and $\| \cdot \|_p$ is the standard $p$-norm, i.e. $\| \varepsilon_t^{Z} \| = \left(\sum_{h=1}^{24} |\varepsilon_{t,h}|^p\right)^{1/p}$, for $p=1,2$. Regardless of the test version, we compute the $p$-value for each model pair and each price area of two one-sided tests: 
\begin{enumerate}
    \item \label{DM null 1} With the null-hypothesis $H_0^1\; :\; \mathbb{E}\left[\Delta^{A,B}\right]\leq0$ for $\Delta^{A,B} \in \left\{\Delta_{t,h}^{A,B}, \Delta_{t}^{A,B}\right\}$,
    \item \label{DM null 2} With the null-hypothesis $H_0^2\; :\; \mathbb{E}\left[\Delta^{A,B}\right]\geq0$ for $\Delta^{A,B} \in \left\{\Delta_{t,h}^{A,B}, \Delta_{t}^{A,B}\right\}$.
\end{enumerate}
In other words, we test for the outperformance of the forecasts of model B by those of model A, and the complementary of the outperformance of the forecasts of model A by those of model B. As per usual, we assume weak stationarity of the loss differential series. The univariate version has the advantage of providing deeper insight into which forecast is better for each hour of the day, while the multivariate version allows for an easier overview of the results as it summarizes the comparison in a single $p$-value. 

\begin{figure}
    \centering
    \includegraphics[width=0.33\linewidth]{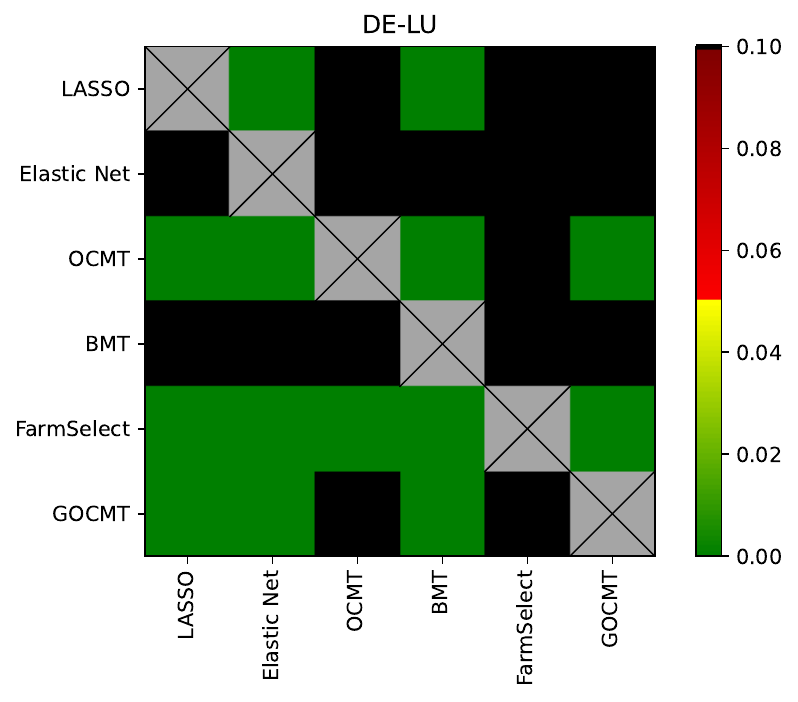}\includegraphics[width=0.33\linewidth]{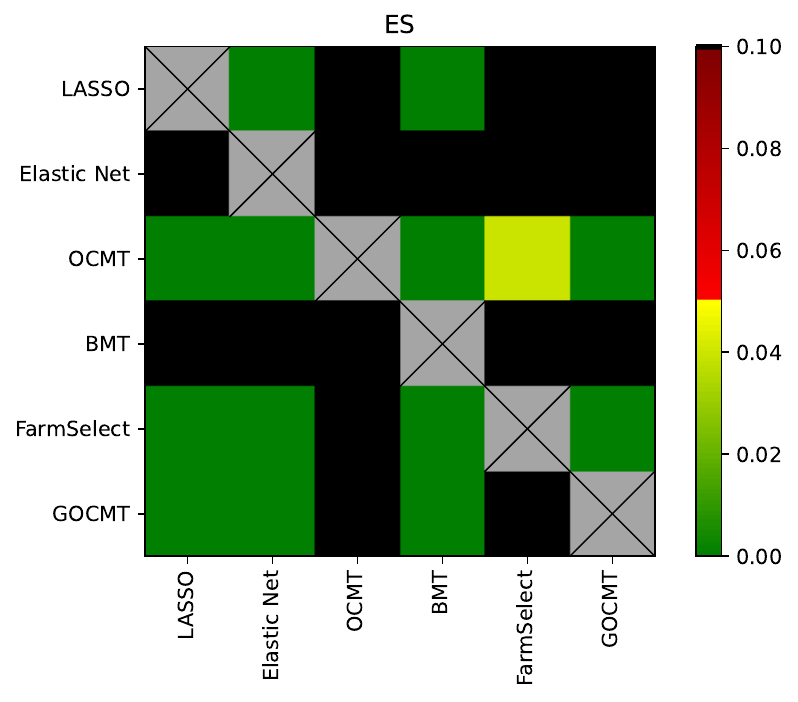}\includegraphics[width=0.33\linewidth]{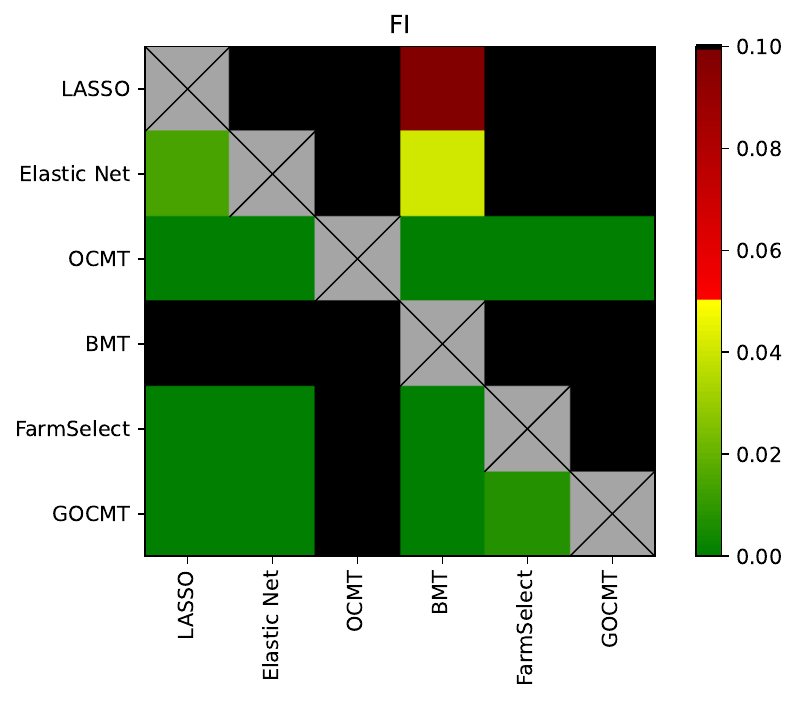}
    \includegraphics[width=0.33\linewidth]{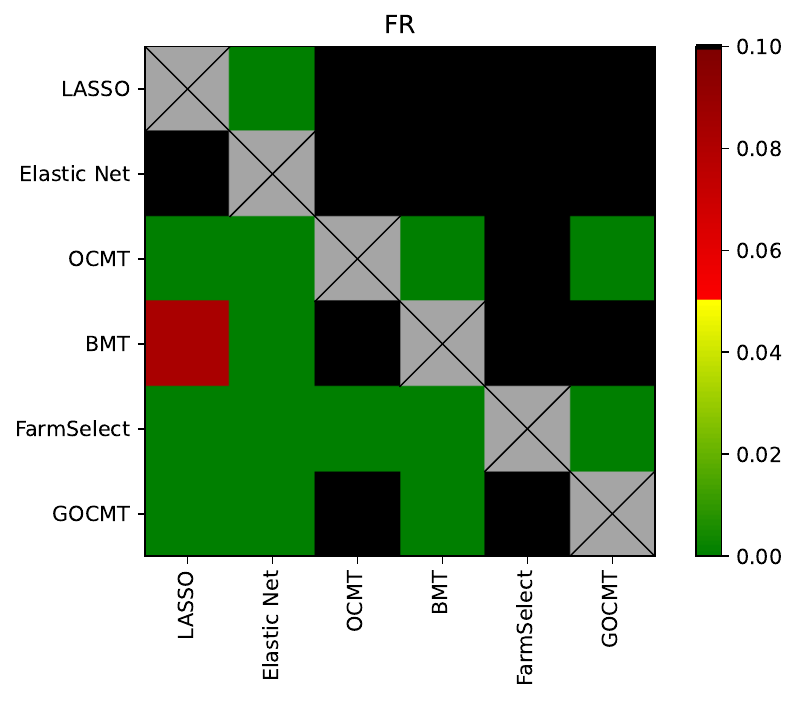}\includegraphics[width=0.33\linewidth]{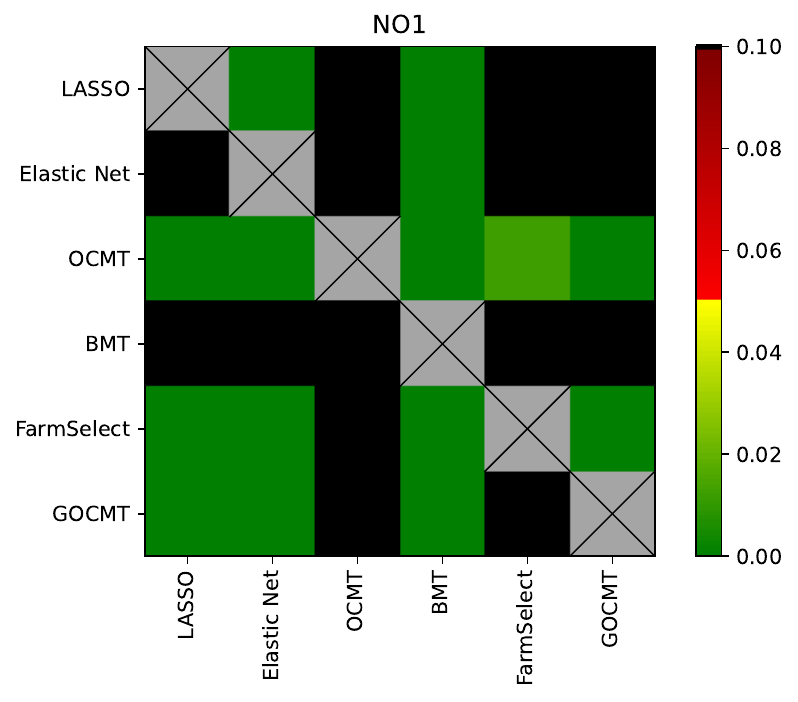}\includegraphics[width=0.33\linewidth]{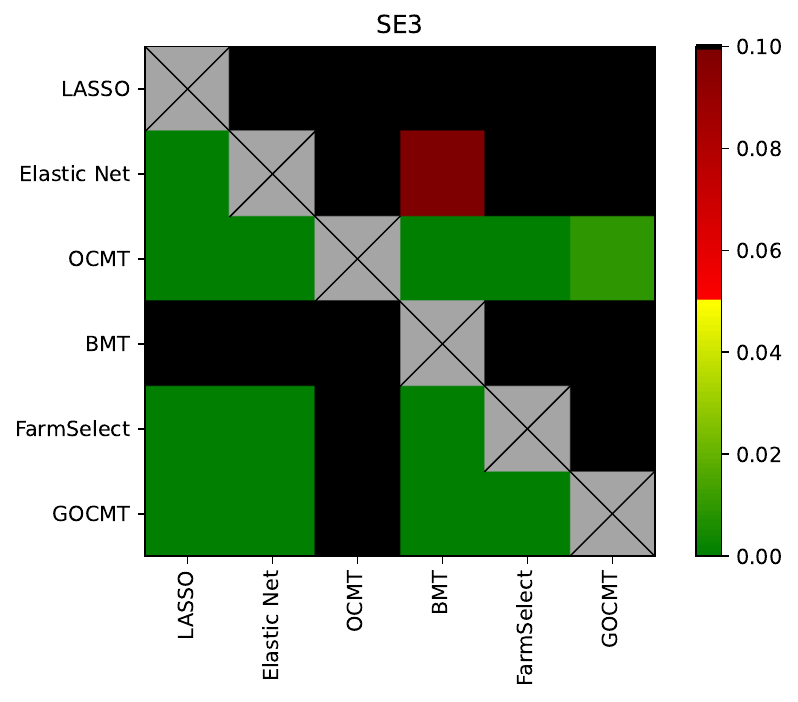}
    \caption{Results of multivariate DM test using the $\ell_1$ error. The heat map displays the range of $p$-values obtained for each of the six markets. Values closer to zero, shown in dark green, indicate stronger evidence that the model on the $x$-axis produces more accurate forecasts than the model on the $y$-axis. Black cells denote $p$-values at or above the upper limit of the color scale, corresponding to $p$-values greater than $0.10$.}
    \label{fig: DM test multivariate}
\end{figure}

Following the suggestion of \citet{lago2021}, we perform the multivariate DM test using the $\ell_1$ norm, i.e., $p=1$, the results of which can be seen in Figure~\ref{fig: DM test multivariate}. 

Overall, the multivariate DM tests support the rankings reported in Table~\ref{tab: hourly metrics}. In DE-LU and ES, neither forecasts generated by BMT and Elastic Net are statistically better than those generated by the other, but both methods produce statistically significantly more accurate forecasts than LASSO, OCMT, FarmSelect, and GOCMT at a 5\% level. In turn, LASSO significantly outperforms each of OCMT, FarmSelect, and GOCMT at a 5\% level. Thus, the tests identify BMT and Elastic Net as the two statistically strongest methods in these markets, with LASSO ranking behind them. In FR, the forecasts generated by Elastic Net are significantly more accurate than those of every other method at a 5\% level, while LASSO outperforms BMT at a 10\% level with no statistical evidence of the opposite being true.

Again, the evidence is particularly favorable to BMT in the Nordic price areas, where in FI, the forecasts generated by BMT are significantly more accurate than those of OCMT, GOCMT, FarmSelect, and Elastic Net at a 5\% level, and LASSO at a 10\% level. LASSO significantly outperforms Elastic Net at a 5\% level, with both methods  significantly outperforming OCMT, FarmSelect, and GOCMT. In NO1, the tests produce a clear ordering among the three leading methods: BMT significantly outperforms Elastic Net, which in turn significantly outperforms LASSO, all at a 5\% level, with all three methods also producing significantly more accurate forecasts than OCMT, FarmSelect, and GOCMT. In SE3, BMT and LASSO remain the two strongest methods, with BMT producing significantly more accurate forecasts than OCMT, and FarmSelect at a 5\% level, and Elastic Net at a 10\% level. Similarly, LASSO significantly outperforms Elastic Net, OCMT, GOCMT and FarmSelect at a 5\% level. No statistical evidence supports either of LASSO and BMT producing more accurate forecasts than the other.

More generally, OCMT, FarmSelect, and GOCMT do not generate significantly more accurate forecasts than BMT, LASSO, or Elastic Net in any of the six price areas. Conversely, the latter three methods significantly outperform OCMT, FarmSelect, and GOCMT at the 5\% level throughout. Most importantly, BMT is significantly outperformed only by Elastic Net and LASSO in FR. Its forecasts are statistically indistinguishable from those of Elastic Net in DE-LU and ES and significantly more accurate in all three Nordic price areas at a 10\% level. Relative to LASSO, BMT is significantly more accurate at the 5\% level in DE-LU, ES, and NO1, at a 10\% level in FI, and statistically indistinguishable in FR and SE3.

\begin{figure}[tb!]
    \centering
    \includegraphics[width=0.33\linewidth]{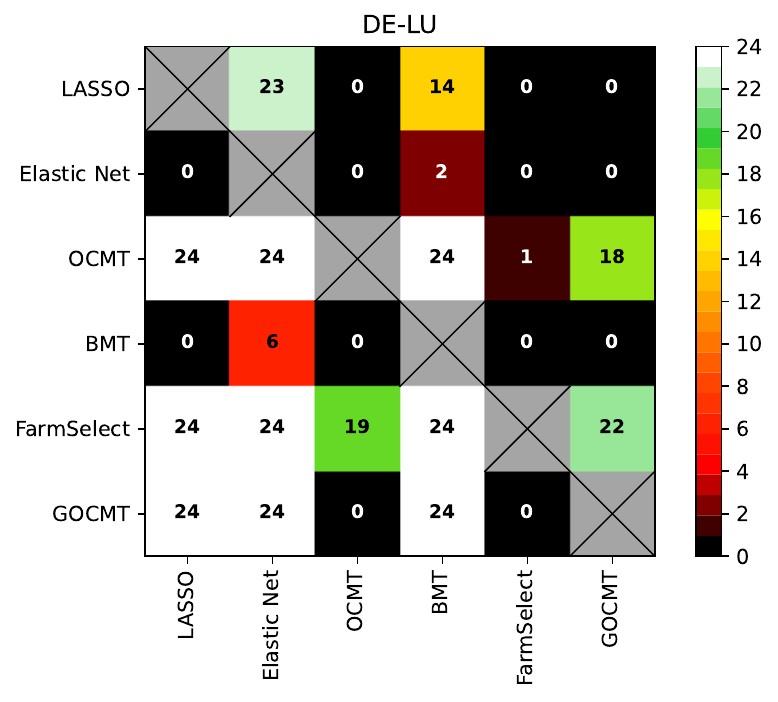}\includegraphics[width=0.33\linewidth]{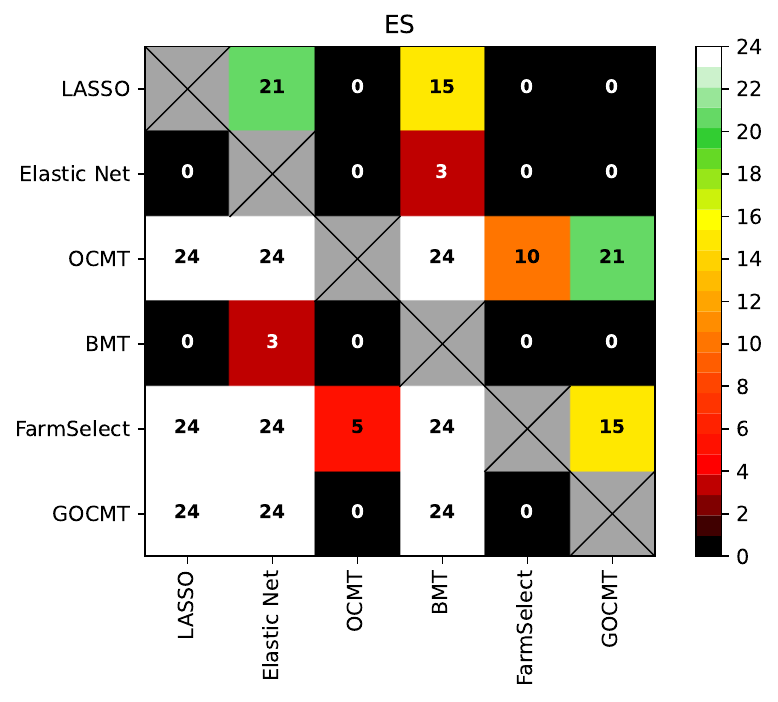}\includegraphics[width=0.33\linewidth]{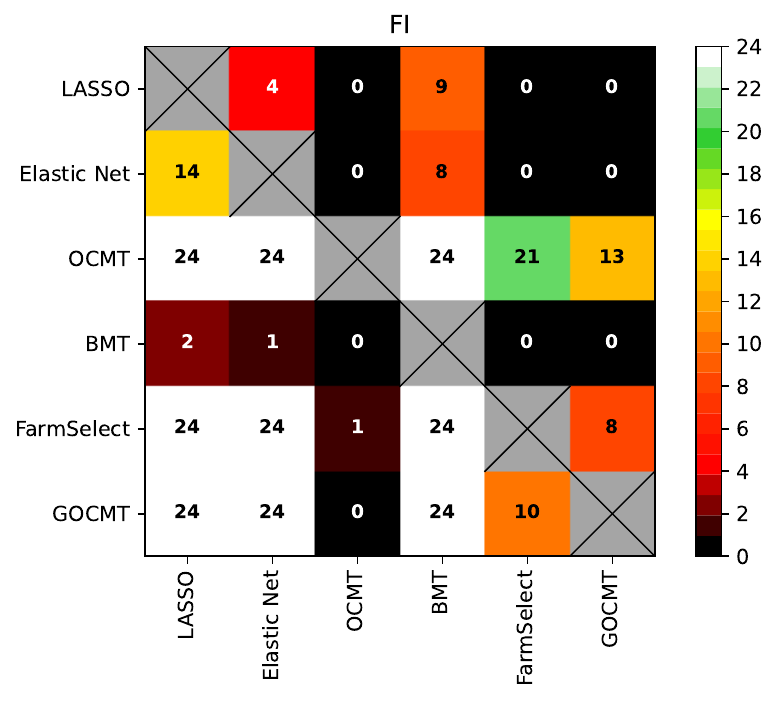}
    \includegraphics[width=0.33\linewidth]{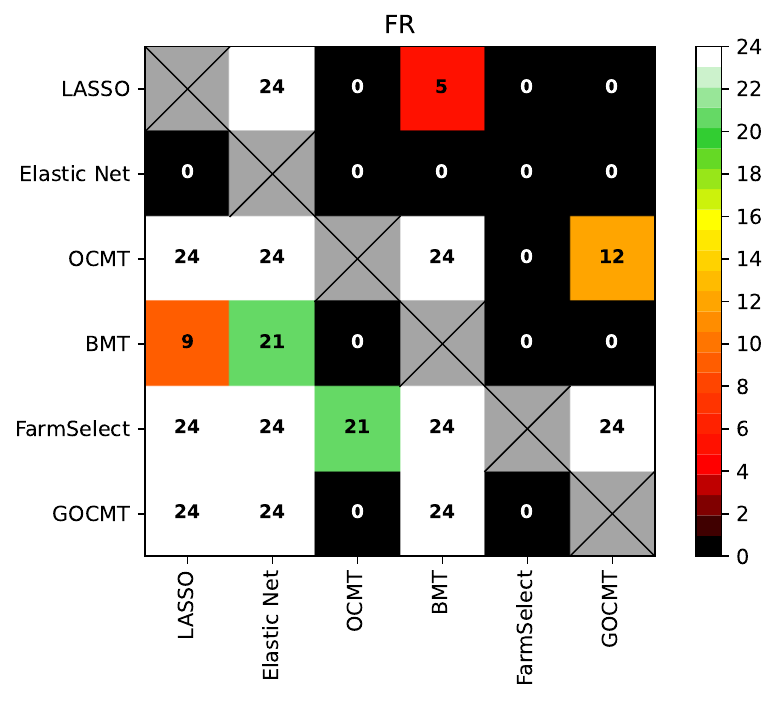}\includegraphics[width=0.33\linewidth]{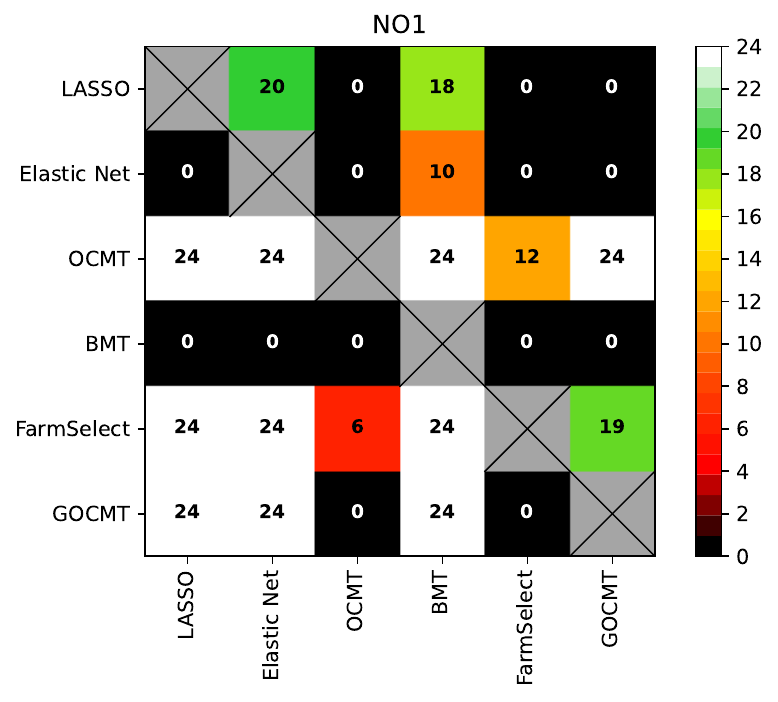}\includegraphics[width=0.33\linewidth]{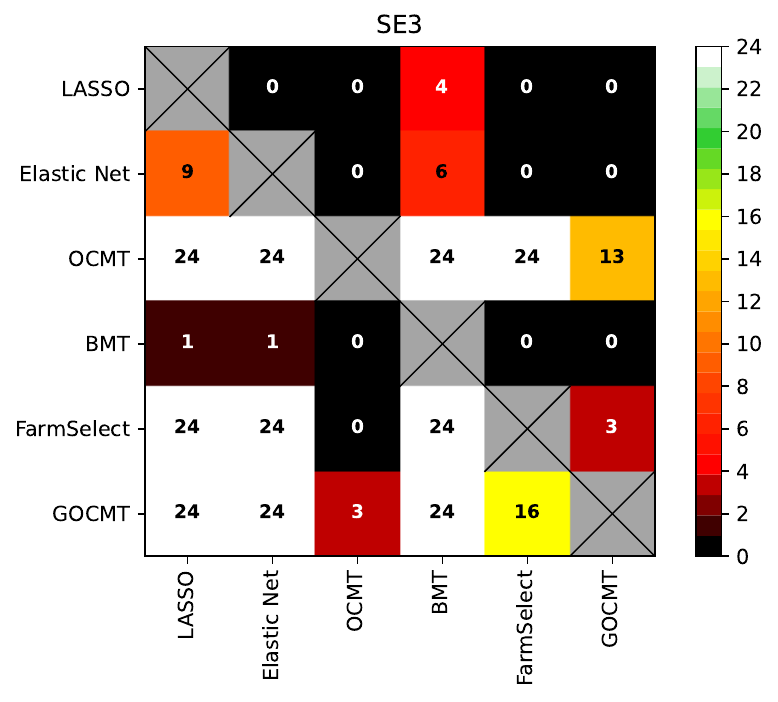}
    \caption{Results of univariate DM test at a 5\% level using the $\ell_1$ error. The heat map displays the number of significant differences in forecasting performance across the 24 hours of the day. A white square indicates that the forecasts of the model on the $x$-axis are significantly better than those of the model on the $y$-axis for all 24 hours. A black square indicates that the forecasts of the model on the $x$-axis are not significantly better than those of the model on the $y$-axis for any of the 24 hours.}
    \label{fig: DM test univariate}
\end{figure}

Results for the univariate DM tests are displayed in Figure~\ref{fig: DM test univariate} for the six price areas. 
In DE-LU, ES, and FR, Elastic Net generates significantly more accurate forecasts than LASSO for 23, 21, and 24 hours, respectively, whereas LASSO outperforms Elastic Net for no hours in these areas. The comparisons between Elastic Net and BMT are more market-dependent. In DE-LU, BMT significantly outperforms Elastic Net for two hours, while Elastic Net outperforms BMT for six hours. In ES, the both methods outperform the other for three hours. These relatively small numbers are consistent with the multivariate tests, which do not identify a significant difference between BMT and Elastic Net in either market. In FR, by contrast, Elastic Net significantly outperforms BMT for 21 hours, whereas BMT is not significantly more accurate for any hours, supporting the results of the multivariate tests and evaluation metrics in Table~\ref{tab: hourly metrics}.

Relative to LASSO, BMT produces significantly more accurate forecasts for 14 hours in DE-LU and 15 hours in ES, while LASSO does not outperform BMT for any hours. The comparison is more balanced in FR, where BMT significantly outperforms LASSO for five hours, and LASSO outperforms BMT for nine hours.

The hourly evidence shifts more clearly in favor of BMT in the Nordic price areas. In FI, BMT generates significantly more accurate forecasts than LASSO and Elastic Net for nine and eight hours, respectively, whereas the reverse occurs for only two and one hour. Comparing LASSO and Elastic Net, the Elastic Net outperforms LASSO for four hours, while LASSO outperforms Elastic Net for 14 hours. In NO1, BMT significantly outperforms LASSO for 18 hours and Elastic Net for ten hours, while neither method significantly outperforms BMT for any hours. Elastic Net also outperforms LASSO for 20 hours, compared with zero hours in the opposite direction, producing the same ordering of BMT, Elastic Net, and LASSO as the multivariate test. BMT also performs strongly in SE3, producing significantly more accurate forecasts than LASSO and Elastic Net for four and six hours, respectively, while in the reverse comparisons, both LASSO and Elastic Net outperform BMT for only one hour. LASSO, in turn, significantly outperforms Elastic Net for nine hours, whereas Elastic Net outperforms LASSO for no hours. Thus, the hourly results support the multivariate finding that BMT and LASSO are the two strongest methods in SE3.

The most uniform results concern the comparisons with OCMT and GOCMT, where each of LASSO, Elastic Net, and BMT significantly outperforms OCMT and GOCMT for the majority of all hours across the six price areas, while neither OCMT nor GOCMT produces significantly more accurate forecasts than any of LASSO, Elastic Net, or BMT for a single hour in any price area. For FarmSelect, the picture is similar.

Overall, the univariate results complement the multivariate test results. Elastic Net has the clearest hourly advantage in FR and also performs strongly relative to LASSO in DE-LU and ES. The direct comparison with BMT is more balanced in DE-LU and favors BMT in ES. In the Nordic price areas, the hourly comparisons consistently favor BMT, particularly in NO1, where neither LASSO nor Elastic Net significantly outperforms BMT for any hours.

\subsection{Computational time}
Since the forecasting models are recalibrated each day, computational time is an important practical consideration. As emphasized by \citet{lago2021}, a method may be unsuitable for operational forecasting if its estimation cannot be completed within the available time window; a marginal improvement in forecast accuracy may not justify substantially greater computational costs. Table~\ref{tab: computational times} therefore reports the minimum and maximum time required for one complete daily recalibration observed over the evaluation period. All computations were performed in Python 3.12 on a dual-socket system equipped with two Intel Xeon Gold 6132 processors operating at 2.60 GHz, providing 28 physical cores and 56 logical processors in total. All methods were executed using a single CPU thread. 

Overall, every method can be recalibrated well within a practically relevant daily forecasting window. Even the longest observed recalibration takes 2 minutes, which is considerably shorter than the 30-minute to one-hour limit discussed by \citet{lago2021}. Nevertheless, the BMT and OCMT-type methods do a full recalibration in 1--3 and 3--7 seconds, respectively, offering a computational time advantage of roughly 5--20 times that of LASSO, Elastic Net, and FarmSelect.

Although this difference is unlikely to be decisive to practitioners for a single daily recalibration, it becomes more consequential in backtest settings. In practice, such backtests may be repeated frequently if new candidate variables, alternative data sources, or new strategies are considered. Depending on the length of the out-of-sample period, BMT can complete a backtest within minutes, whereas the penalised-regression methods may require several hours or, in some cases, days. The computational advantage of BMT is thus particularly valuable to practitioners during model development and robustness analysis.

\begin{table}[h!]
\centering
\caption{Computational times that each method requires to perform a daily recalibration for the six considered price areas.}
\label{tab: computational times}
\begin{tabular*}{\textwidth}{@{\extracolsep{\fill}}lcccccc}
\hline \hline
 & LASSO & Elastic Net & OCMT & BMT & FarmSelect & GOCMT \\
\midrule
DE-LU & 50--115 s & 15--27 s & 3--6 s & 2--3 s & 52--119 s & 3--4 s \\
ES & 54--116 s & 19--46 s & 3--6 s & 2--2 s & 57--133 s & 2--4 s \\
FI & 58--127 s & 45--125 s & 3--6 s & 2--3 s & 53--117 s & 2--3 s \\
FR & 49--110 s & 15--74 s & 3--7 s & 2--3 s & 47--107 s & 3--5 s \\
NO1 & 40--97 s & 17--81 s & 3--6 s & 1--2 s & 41--93 s & 2--6 s \\
SE3 & 58--123 s & 44--95 s & 3--7 s & 2--2 s & 51--117 s & 3--4 s \\
\hline \hline
\end{tabular*}
\end{table}
\section{Conclusion}\label{sec:conclusion}
We have considered the problem of variable selection for forecasting electricity spot prices in a high-dimensional setting. We have compared the performance of six variable selection procedures, two traditional regularization methods, LASSO and Elastic Net, and four screening-based alternatives, FarmSelect, OCMT, GOCMT, and the recently proposed Boosting Multiple Testing (BMT) procedure, none of which had previously been applied in the electricity price forecasting context. Using extensive spot price datasets from six major European electricity markets, we find that LASSO and Elastic Net achieve similar forecasting performance and outperform  FarmSelect, OCMT, and GOCMT on average. BMT, by contrast, achieves statistically comparable accuracy to these shrinkage benchmarks while selecting a substantially smaller number of variables, using less than one-tenth of those retained by LASSO and Elastic Net, and at a considerably lower computational cost.

These findings suggest that the over-parameterization typically associated with regularization methods is not a necessary price for predictive accuracy in electricity price forecasting. By achieving comparable forecasting performance with a far more parsimonious specification, BMT offers a viable alternative to shrinkage-based methods, one that yields models which are both more interpretable and less costly to estimate. This finding demonstrates that model transparency and computational efficiency, often as important as predictive accuracy in practice, need not be sacrificed for forecasting performance. Given the growing complexity and dimensionality of candidate predictor sets in electricity markets, driven by the increasing integration of renewable generation and cross-border interconnections, screening-based procedures of this kind warrant further attention as a practical complement to established regularization approaches in applied forecasting.

\section*{Acknowledgements}

We thank Lasse Bork for valuable comments and suggestions that helped improve the quality of the paper. Mikkel Mandrup acknowledges support from Innovation Fund Denmark, grant number 4365-00024B.

\bibliography{references}

\newpage
\appendix

\section{Penalised-regression approaches for variable selection}\label{appx_penalised_reggressions}

In what follows, we describe the penalised-regression methods used in our empirical analysis. These methods have been proven to be highly effective when the number of candidate regressors is very large. 

\subsection{LASSO}\label{appx_lasso}
The Least Absolute Shrinkage and Selection Operator (LASSO) introduced in \citet{lasso} uses an $\ell_1$ penalty factor to introduce shrinkage and variable selection. First introduced in the context of EPF in \citet{Uniejewski2016} under the name LassoX, they conclude that LassoX outperforms other variable selection methods, such as stepwise regression. Other uses of the LASSO-type models in EPF for day-ahead spot price forecasting include \citet{ZielWeron2018} and \citet{lago2021}, where it was introduced under the names 24Lasso$_{DoW,nl}$ and LASSO-Estimated AutoRegressive (LEAR) model, respectively. Throughout this study, we will simply refer to the method as LASSO. The minimization problem solved in LASSO is the following
\begin{equation}\label{eq: lasso}
    \hat\beta = \underset{\beta \in \R^{n+1}}{\arg \min} \left\{ \left( y_t - \beta_0 - \sum_{i=1}^d \beta_i x_{t,i} \right)^2 + \lambda \sum_{i=1}^d \abs{\beta_i} \right\},
\end{equation}
where $\lambda \geq 0$ is a tuning parameter to be decided by the user. 

\subsection{Elastic Net}\label{appx_elasticnet}
Introducing both an $\ell_1$ and an $\ell_2$ penalty, elastic net selects variables like LASSO and shrinks together the coefficients of correlated variables like ridge regression. While not as widely used as LASSO, elastic net was first used by \citet{Uniejewski2016} in the context of EPF, where it outperformed all considered variable selection methods, including LASSO, in terms of forecast accuracy. Other uses include \citet{MUNIAIN2020} for forecasting day-ahead prices in Germany, and \citet{AGRAWAL2019} for real-time predictions in the New England electricity market. The minimization problem to be solved in elastic net takes the form:
\begin{equation}\label{eq: elastic net}
    \hat\beta = \underset{\beta \in \R^{n+1}}{\arg \min} \left\{ \left( y_t - \beta_0 - \sum_{i=1}^d \beta_i x_{t,i} \right)^2 + \lambda \left( \frac{1-\alpha}{2}\sum_{i=1}^d \abs{\beta_i} + \alpha\sum_{i=1}^d \beta_i^2 \right)\right\}, 
\end{equation}
where $\alpha \in [0,1]$ and $\lambda\geq0$ are tuning parameters. For $\alpha = 1$, elastic net reduces to the LASSO, while for $\alpha=0$ it becomes ridge regression. The shrinkage of correlated regressors is an especially useful feature in the context of EPF, where variables can be highly correlated. A disadvantage of elastic net, however, is the fact that it has two tuning parameters $\alpha, \lambda$ compared to just one in LASSO. An approach for determining $\alpha$ and $\lambda$ is to do a double grid search and use an information criterion to determine the optimal values. Following \citet{Uniejewski2016} we do a double grid search over exponentially decreasing $\lambda$'s and three values of $\alpha = 0.25, \alpha = 0.50, \alpha = 0.75$, and choose the pair $\{\lambda, \alpha\}$ based on the AIC. More specifically, for each $\alpha$ we construct the grid of $\lambda$'s as follows: define
$$
\lambda_{max} = \frac{\max_{j \in\{1, \ldots, n\}} |x_j^\top y|}{D\alpha},
$$
and set $\lambda_{min} = 10^{-3}\lambda_{max}$, then the grid is constructed as
$$
\lambda_m = \lambda_{max}\left( \frac{\lambda_{min}}{\lambda_{max}} \right)^{m/(M-1)}, \quad m = 0, 1, \ldots, M - 1.
$$

\subsection{Factor-Adjusted Regularized Model Selection}\label{appx_farmselect}
The Factor-Adjusted Regularized Model Selection (FarmSelect), proposed by \citet{FarmSelect}, is designed for high-dimensional regression problems in which the candidate covariates exhibit strong dependence. FarmSelect addresses this issue by decomposing the candidate variables into a small number of common factors and weakly correlated idiosyncratic components. Variable selection is then performed on the idiosyncratic components, while the estimated common factors are included as unpenalised regressors.

By removing the dependence attributable to a small number of common factors before applying regularisation, FarmSelect transforms the problem from one with highly correlated covariates to one with weakly correlated ones. The numerical results of \citet{FarmSelect} indicate that the method is superior to both LASSO and elastic net in terms of recovering the correct model specification in settings of highly correlated covariates.

Let $X =(x_1,\ldots,x_T)^\top \in \mathbb{R}^{T\times n}$ denote the observation matrix of the $n$ candidate covariates. FarmSelect assumes that the vector of candidate variables follows an approximate factor model of the form
$$
x_t = Bf_t + u_t, \qquad t=1,2,\ldots,T,
$$
where $x_t=(x_{t,1},\ldots,x_{t,d})^\top$, $f_t\in\mathbb{R}^\zeta$ is a vector of $\zeta \in \N$ latent common factors, $B\in\mathbb{R}^{d\times \zeta}$ is a factor loading matrix, and $u_t\in\mathbb{R}^d$ are the idiosyncratic components. In matrix form, the decomposition is
\begin{equation}\label{eq: factor model}
    X = FB^\top + U,
\end{equation}
where $F=(f_1,\ldots,f_T)^\top\in\mathbb{R}^{T\times \zeta}$ and $U=(u_1,\ldots,u_T)^\top\in\mathbb{R}^{T\times d}$. FarmSelect is comprised of a two-step procedure: in the first step, the factor model in \eqref{eq: factor model} is estimated to recover an estimate of the idiosyncratic component. Following this, a penalised regression is performed using the estimated factors and idiosyncratic components in the second step. We specify the FarmSelect using LASSO, but note that it may be implemented for any penalised regression specification.

\paragraph{Step 1: Estimation of approximate factor model}
Since the factors and loadings are unobserved, they are estimated by principal component analysis (PCA). Thus, for a given number of factors $\zeta$, the columns of $\hat F/\sqrt{D} = Q$ are the eigenvectors associated with the $\zeta$ largest eigenvalues of $XX^\top$. By the OLS estimator of $B$ we get
\begin{align*}
    \hat B &= X^\top \hat F(\hat F^\top \hat F)^{-1} \\
    & = X^\top \hat F (DQ^\top Q)^{-1} \\
    &= \frac{1}{D}X^\top \hat F
\end{align*}
from which the estimated idiosyncratic components follow as $\hat U = X-\hat F\hat B^\top$.

To specify the number of factors $\zeta$, \citet{FarmSelect} uses a modified eigenvalue-ratio criterion to select the minimum number such that the idiosyncratic components are weakly correlated. Specifically, $\hat \zeta$ is chosen according to
$$
\hat \zeta = \underset{0\leq \zeta\leq \zeta_{\max}}{\arg\min} \frac{\lambda_{\zeta+1}(XX^\top)+C_D}{\lambda_\zeta(XX^\top)+C_D},
$$
where $\lambda_\zeta(XX^\top)$ denotes the $\zeta$th largest eigenvalue of $XX^\top$, $\zeta_{\max} \in \N$ is a prescribed upper bound, and $C_D\in\R$. In the empirical application, we follow the implementation in the FarmSelect R package and choose $C_D=0$ while limiting ourselves to a maximum of 20 factors, $\zeta_{max}=20$.

\paragraph{Step 2: Factor-adjusted penalised regression}
FarmSelect replaces the original, strongly correlated regressors in $X$ with the estimated factors, $\hat F$, and idiosyncratic components, $\hat U$. While FarmSelect is applicable to generalized linear models and a broad class of penalised regression problems, we restrict attention to linear regression with an $\ell_1$ penalty (LASSO), as this is the setting we will be considering in the empirical application. In this case, the regression model reads:
$$
y_t = \beta_0\mathbf{1}_T+\hat F\gamma+\hat U\beta+\varepsilon_t,
$$
where $\gamma \in \R^{\hat\zeta}$ and $\beta\in \R^{d}$. Thus, by lifting the covariate space from $\R^d$ to $\R^{d+\hat\zeta}$, we replace the highly correlated variables in $X$ with weakly correlated ones. In the estimation, $\gamma$ are treated as nuisance parameter and are not penalised, whereas the coefficients $\beta$ associated with the idiosyncratic components are subject to regularisation. Hence, when using an $\ell_1$ penalty the FarmSelect estimator is
$$
(\hat \beta_0,\hat\gamma,\hat\beta) = \underset{\substack{ \beta_0\in\mathbb{R},\, \gamma\in\mathbb{R}^{\hat \zeta},\, \beta\in\mathbb{R}^n}}{\arg\min} \left\{\left( y_t -\beta_0 - \sum_{i=1}^{\hat\zeta} \gamma_i \hat f_{t,i}- \sum_{i=1}^d \beta_i \hat u_{t,i} \right)^2 + \lambda \sum_{i=1}^d |\beta_i| \right\},
$$
for $\lambda>0$. Equivalently, by solving the least squares problem with respect to $\gamma$ we get the profiled least squares problem
\begin{equation}\label{eq: profiled least squares}
    \hat\beta=\arg\min_{\beta\in\mathbb{R}^{d+1}}\left\{\frac{1}{2T} \left\| \hat M_Zy-\hat M_Z\hat U\beta\right\|_2^2+\lambda\|\beta\|_1 \right\},
\end{equation}
where $\hat M = I_D - \hat P$ and $\hat P = \hat F(\hat F^\top\hat F)^{-1}\hat F^\top$ are the projection matrices onto, and orthogonal to, the space spanned by the estimated factors. Thus, FarmSelect can be interpreted as removing the factor variation from the response variable and the idiosyncratic components and then applying LASSO to this factor-adjusted regression problem.

As for LASSO, we use the LARS LASSO to select $\lambda$ in FarmSelect based on the AIC.

\section{Data description figures}\label{appendix: data figures}
\begin{figure}[!ht]
    \centering
    \includegraphics[width=\linewidth]{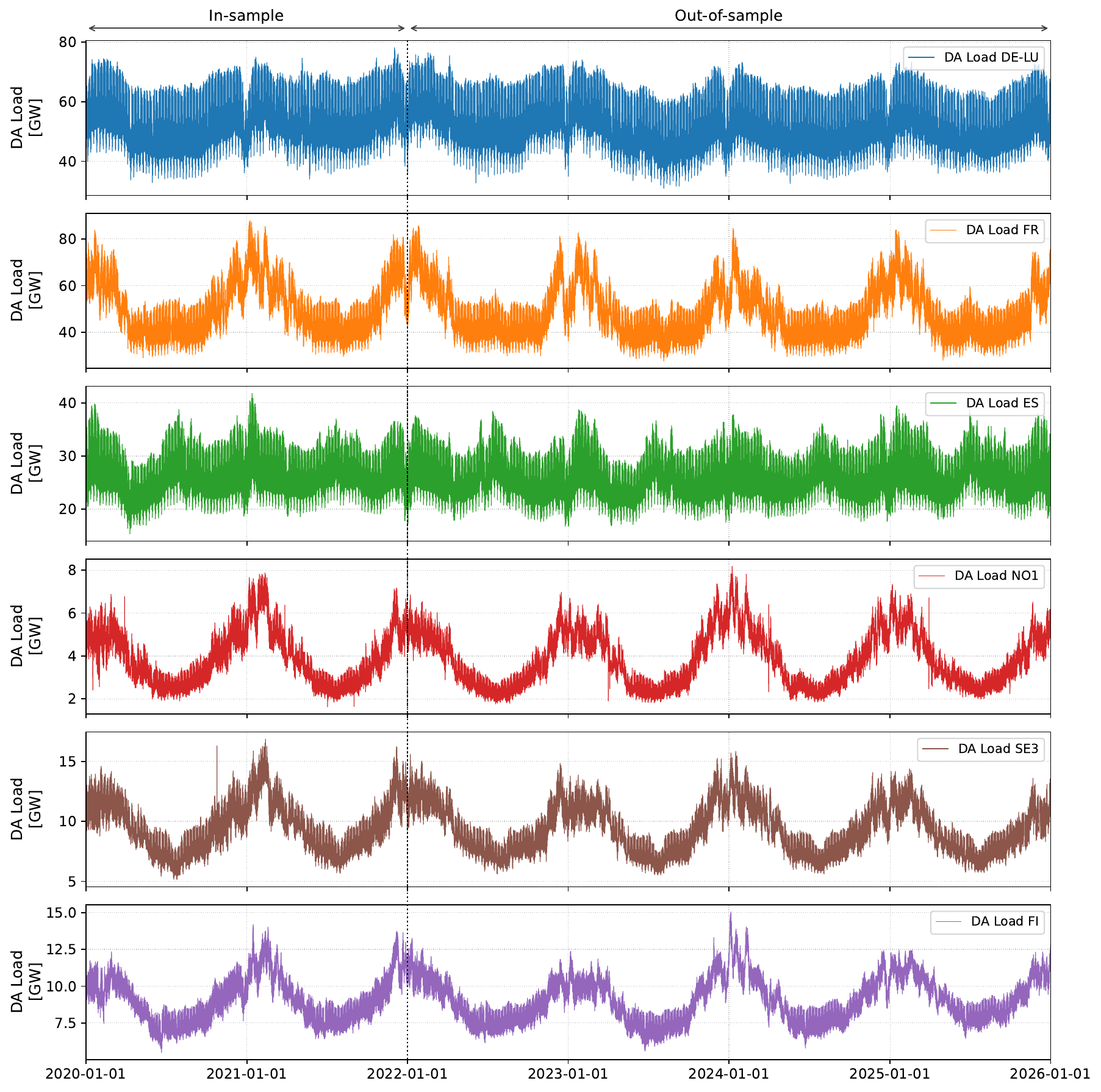}
    \caption{DA load forecasts for the six considered price areas.}
    \label{fig: DA load}
\end{figure}
\begin{figure}[!ht]
    \centering
    \includegraphics[width=\linewidth]{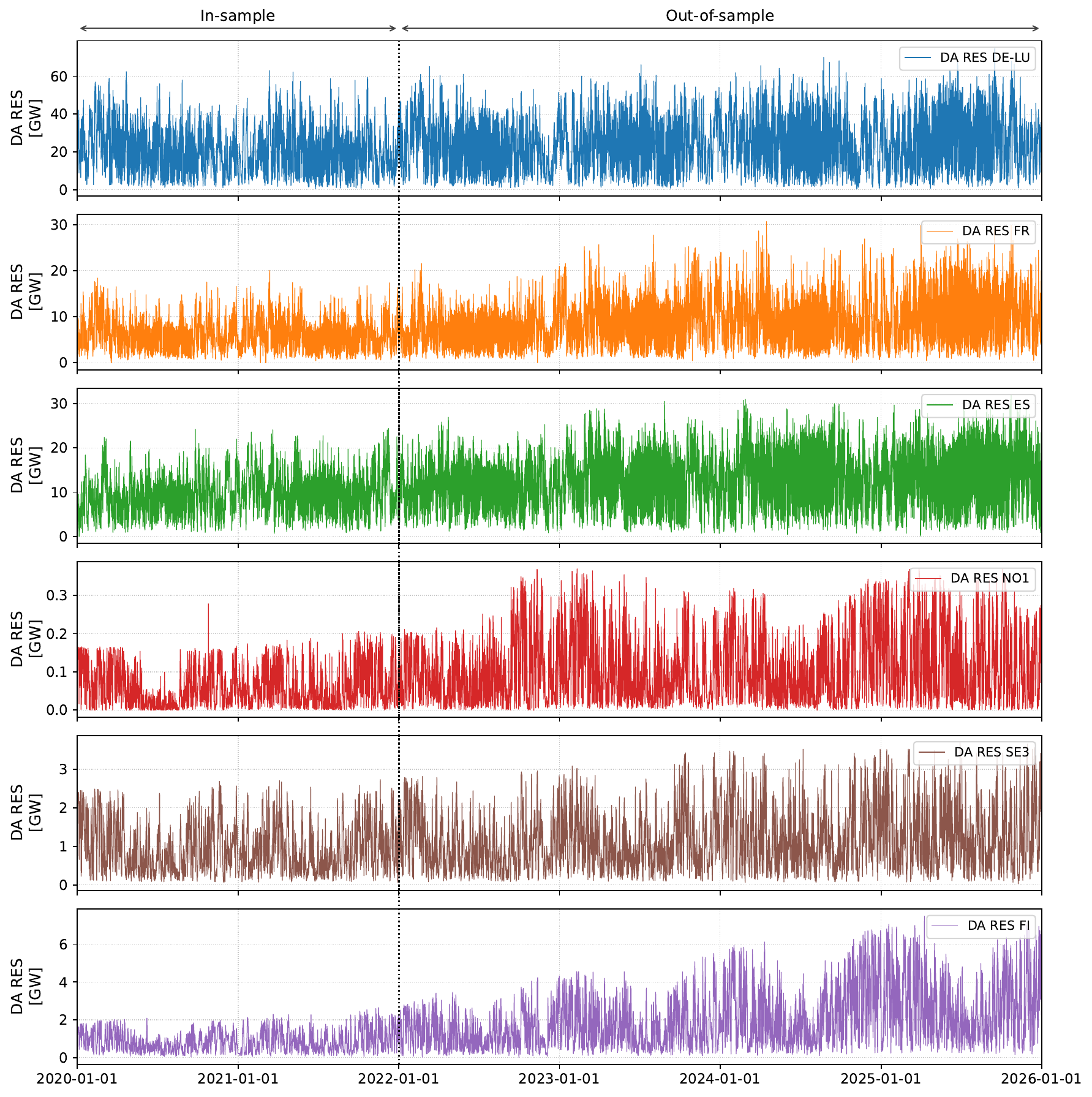}
    \caption{DA RES forecasts for the six considered price areas.}
    \label{fig: DA RES}
\end{figure}

\begin{figure}[H]
    \centering
    \includegraphics[width=\linewidth]{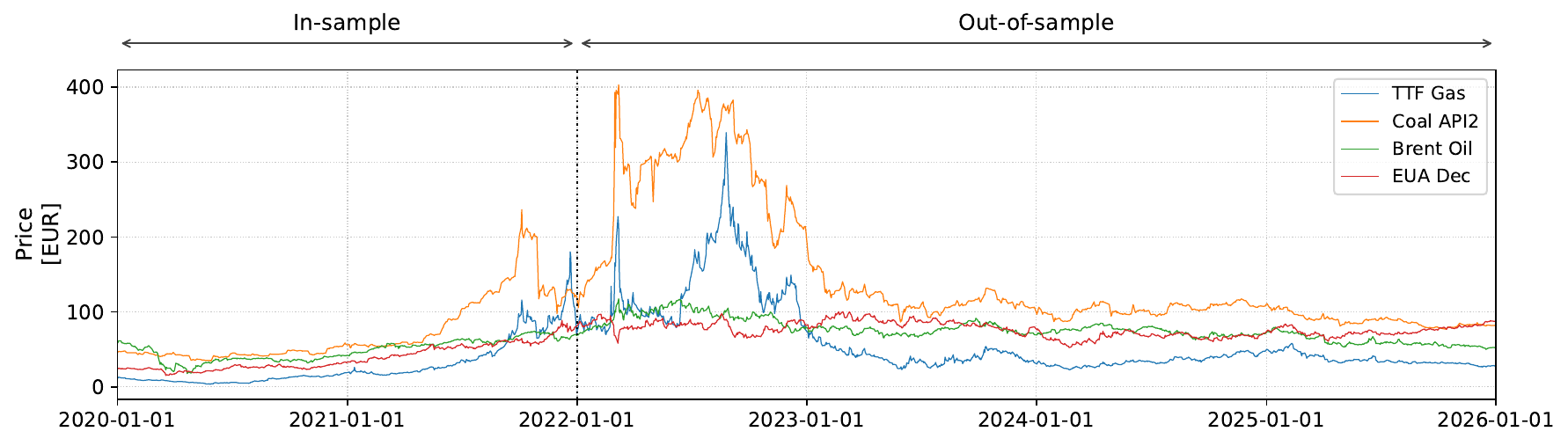}
    \caption{Closing prices for the front-month contracts on TTF, API2, and Brent oil along with EUA Dec closing prices.}
    \label{fig: fuels}
\end{figure}

\section{Result figures}\label{appendix: result figures}

\begin{figure}[!ht]
    \centering
    \includegraphics[width=\linewidth]{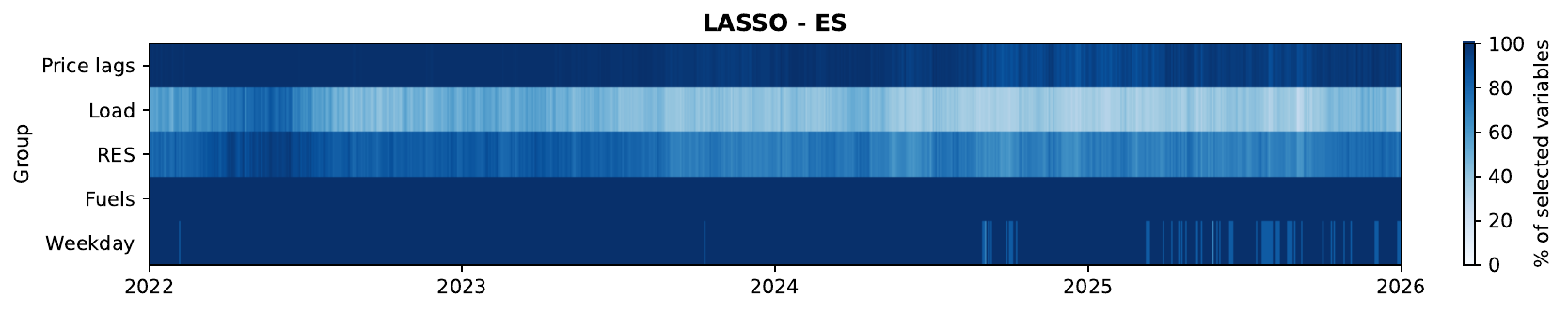}
    \includegraphics[width=\linewidth]{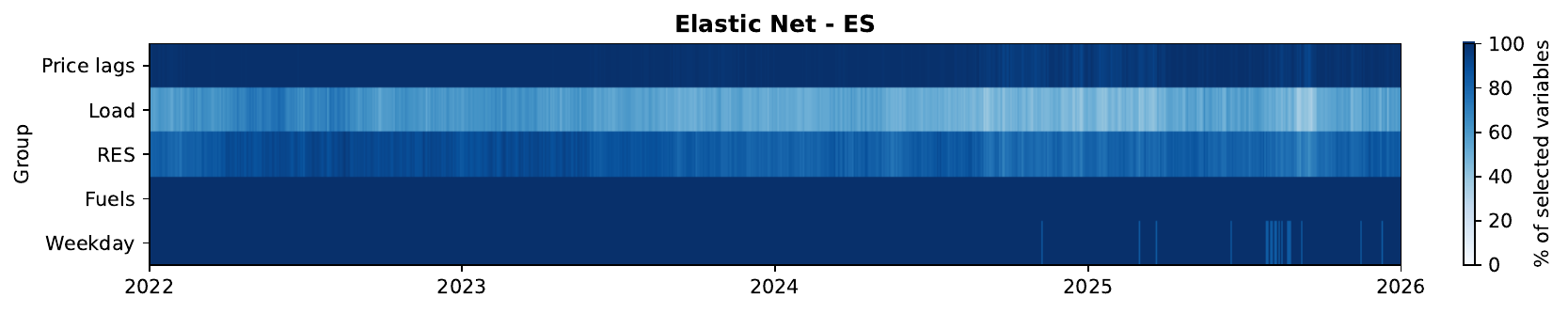}
    \includegraphics[width=\linewidth]{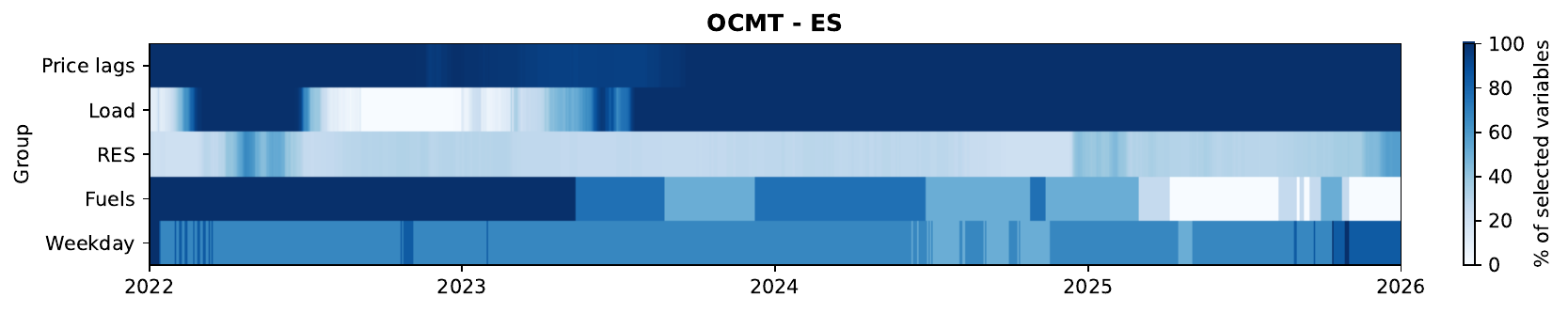}
    \includegraphics[width=\linewidth]{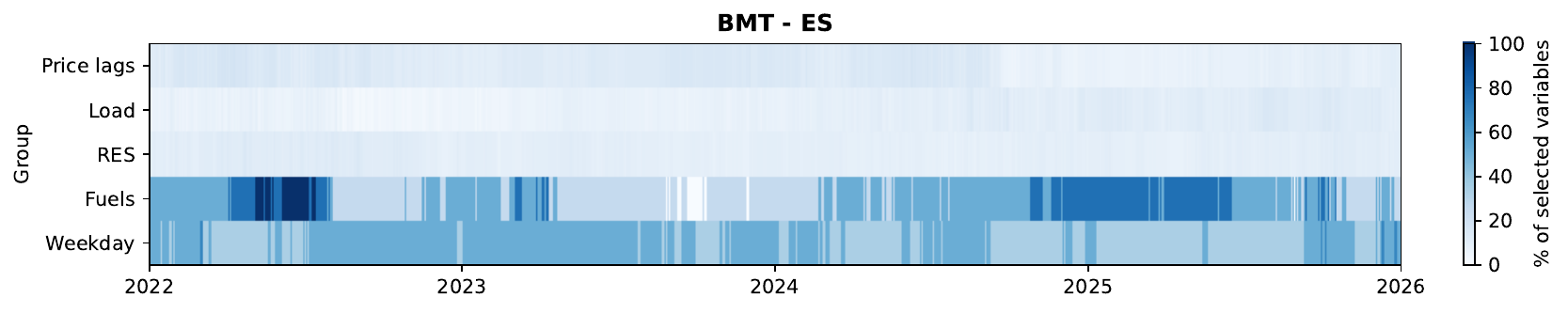}
    \includegraphics[width=\linewidth]{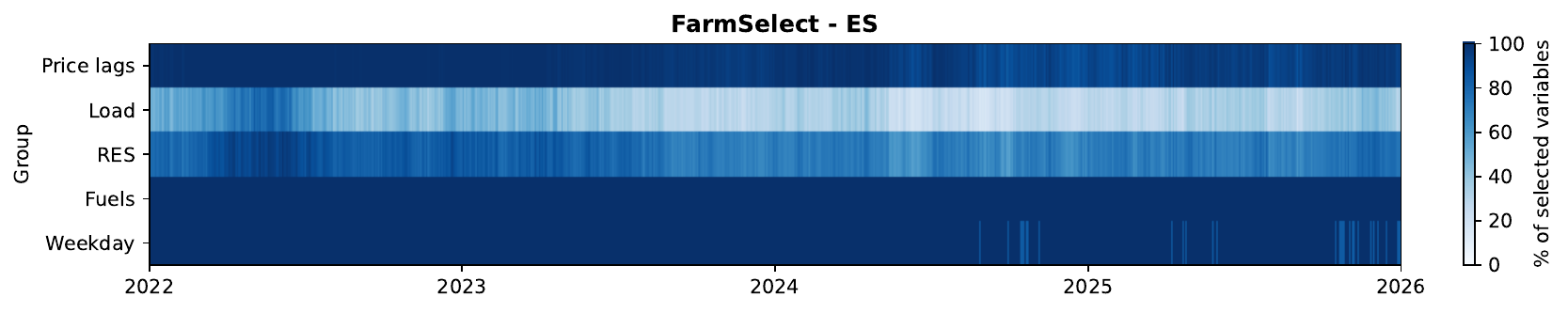}
    \includegraphics[width=\linewidth]{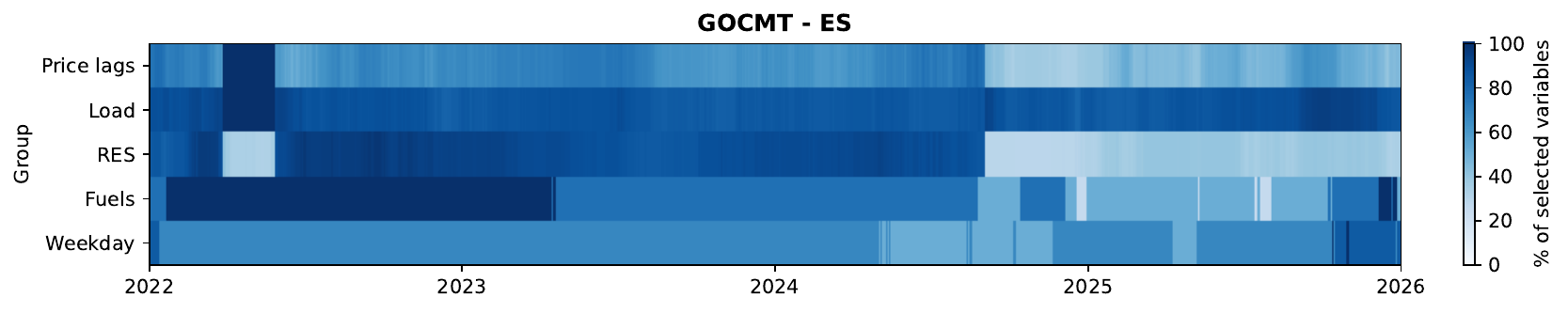}
    \caption{Heat map of the percentage of selected number of variables for each method from the groups Price lags, Load, RES, Fuels, and Weekday.}
\end{figure}

\begin{figure}[!ht]
    \centering
    \includegraphics[width=\linewidth]{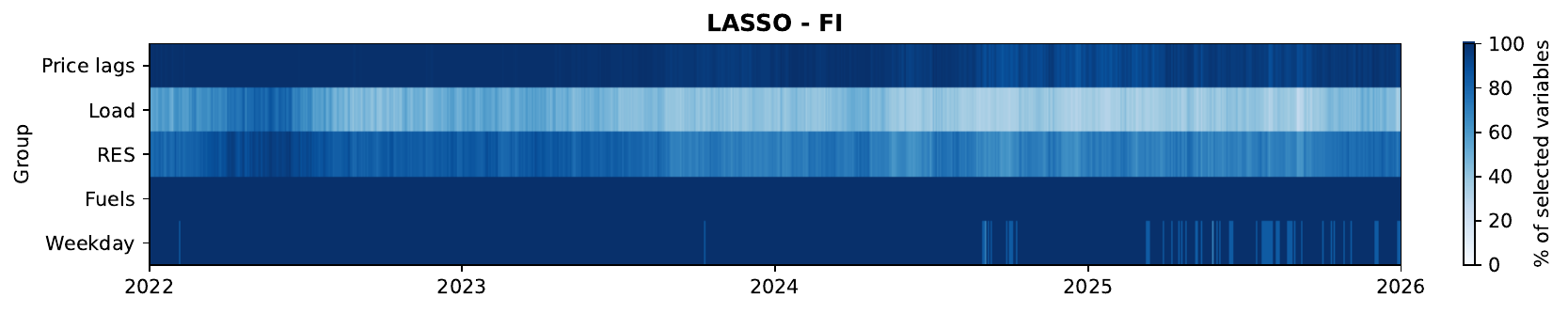}
    \includegraphics[width=\linewidth]{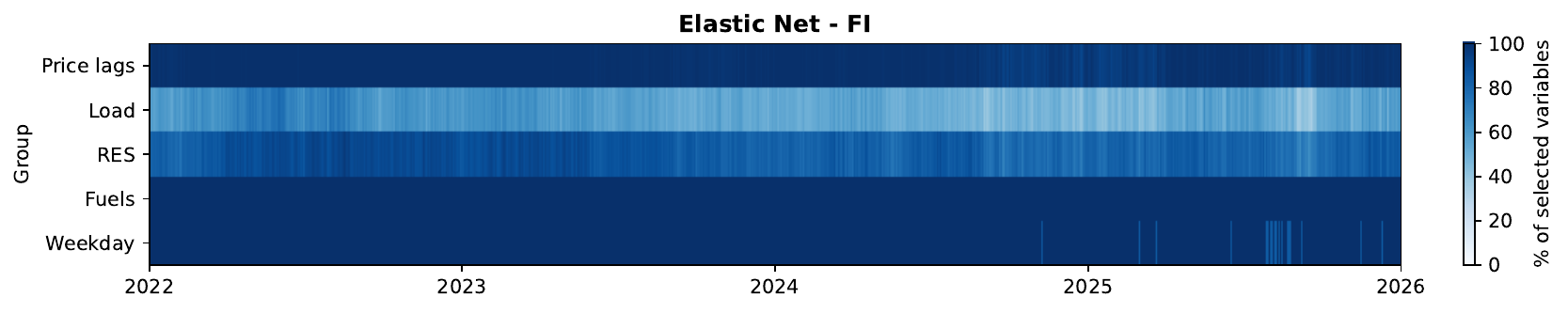}
    \includegraphics[width=\linewidth]{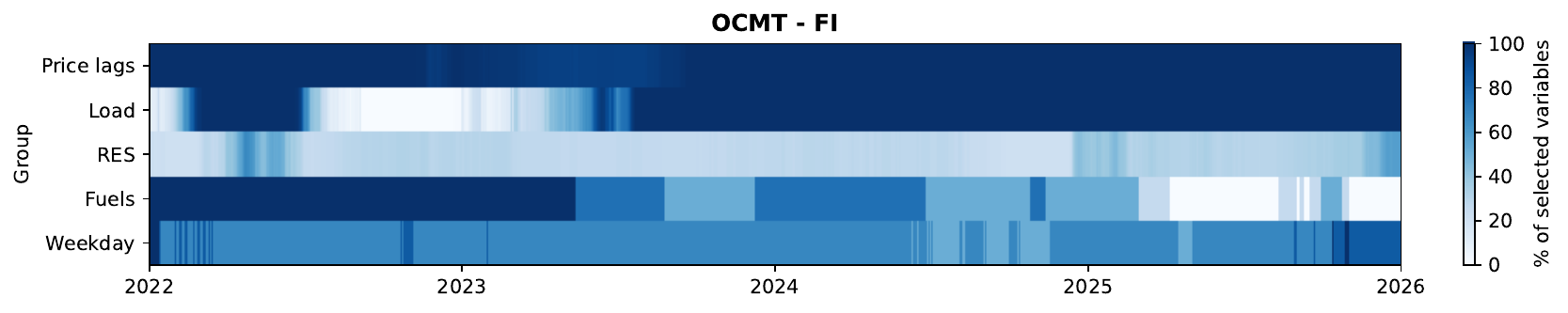}
    \includegraphics[width=\linewidth]{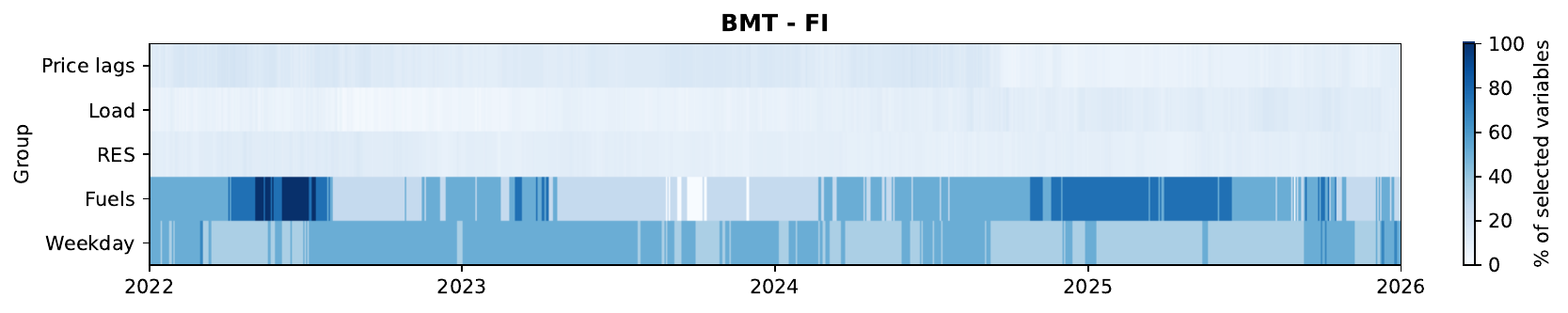}
    \includegraphics[width=\linewidth]{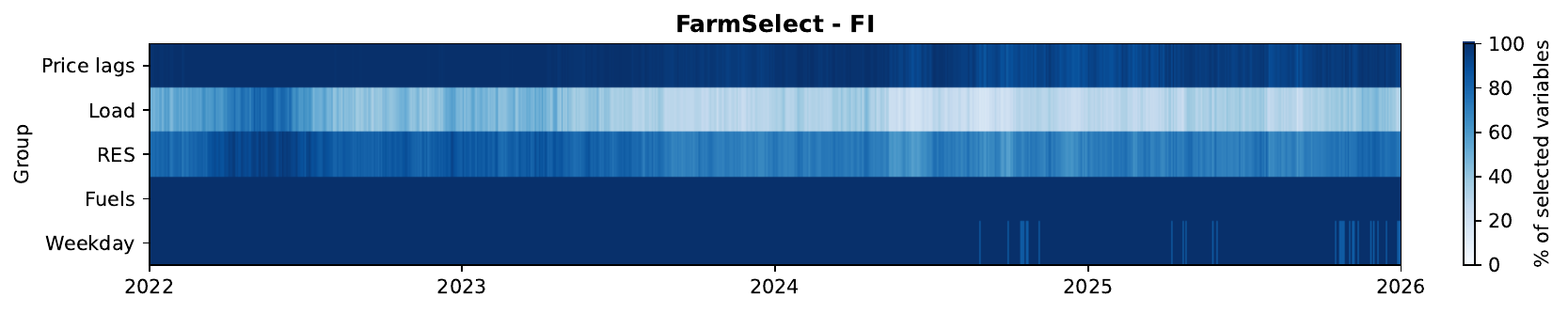}
    \includegraphics[width=\linewidth]{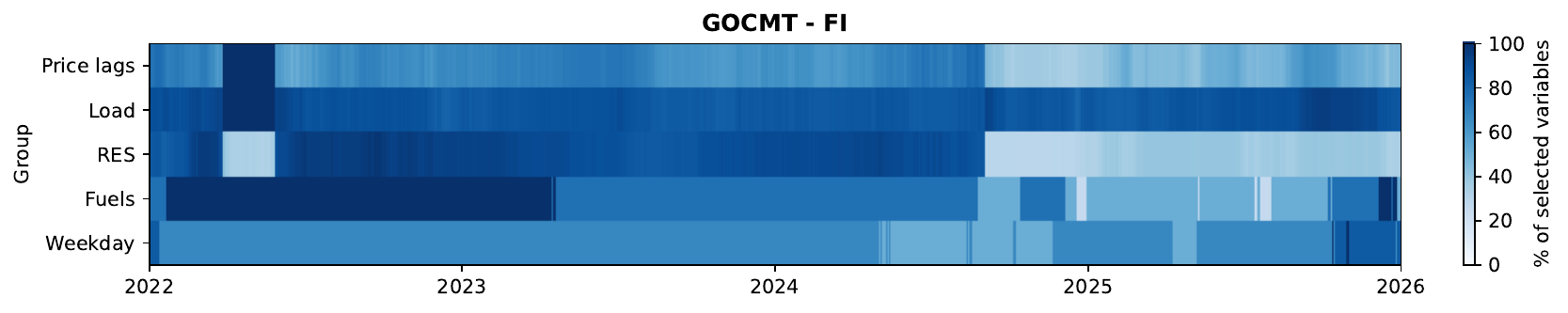}
    \caption{Heat map of the percentage of selected number of variables for each method from the groups Price lags, Load, RES, Fuels, and Weekday.}
\end{figure}

\begin{figure}[!ht]
    \centering
    \includegraphics[width=\linewidth]{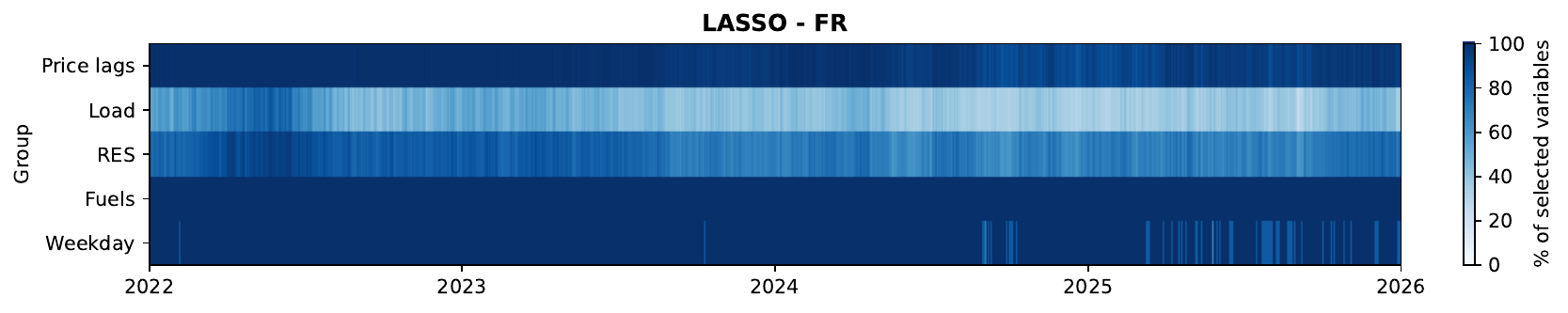}
    \includegraphics[width=\linewidth]{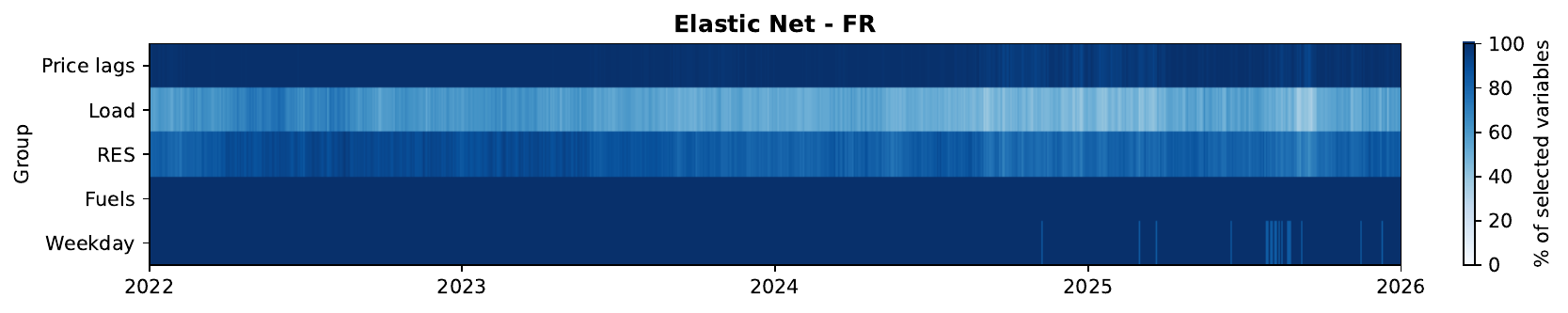}
    \includegraphics[width=\linewidth]{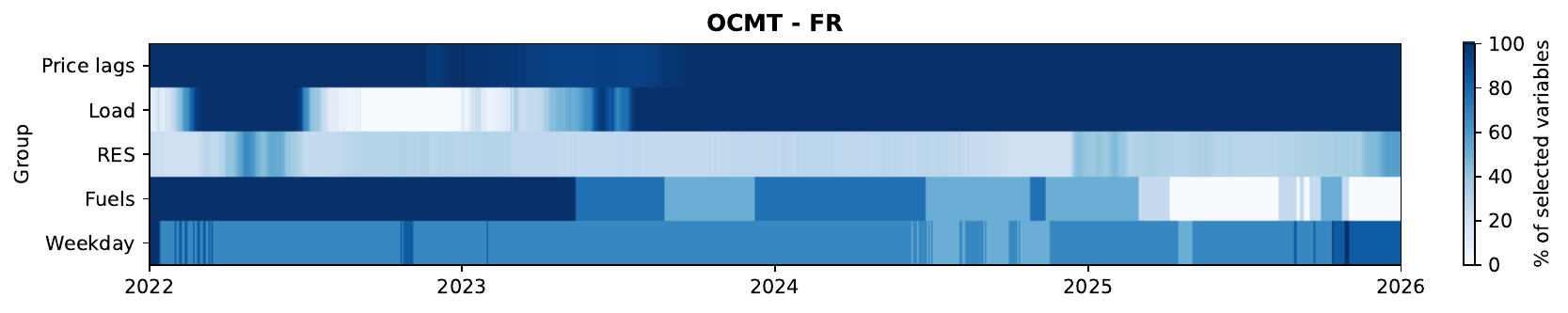}
    \includegraphics[width=\linewidth]{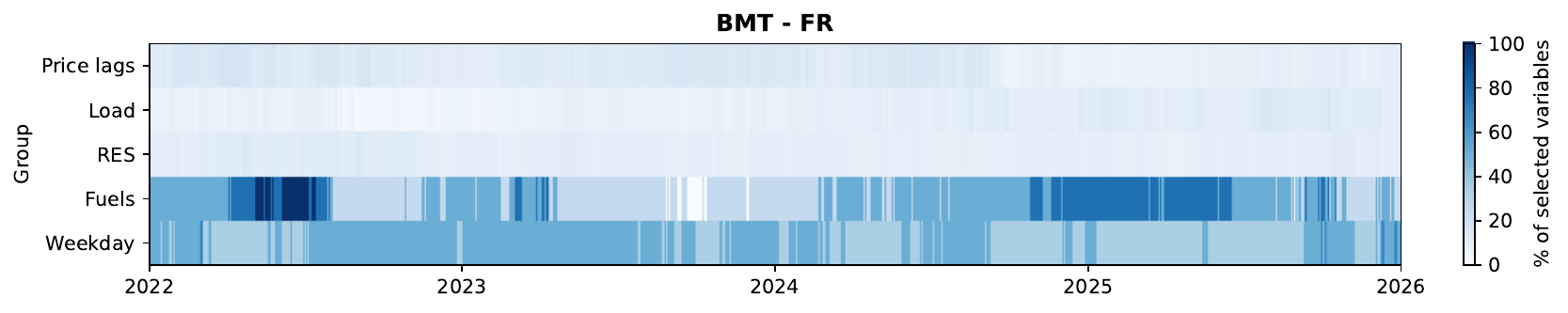}
    \includegraphics[width=\linewidth]{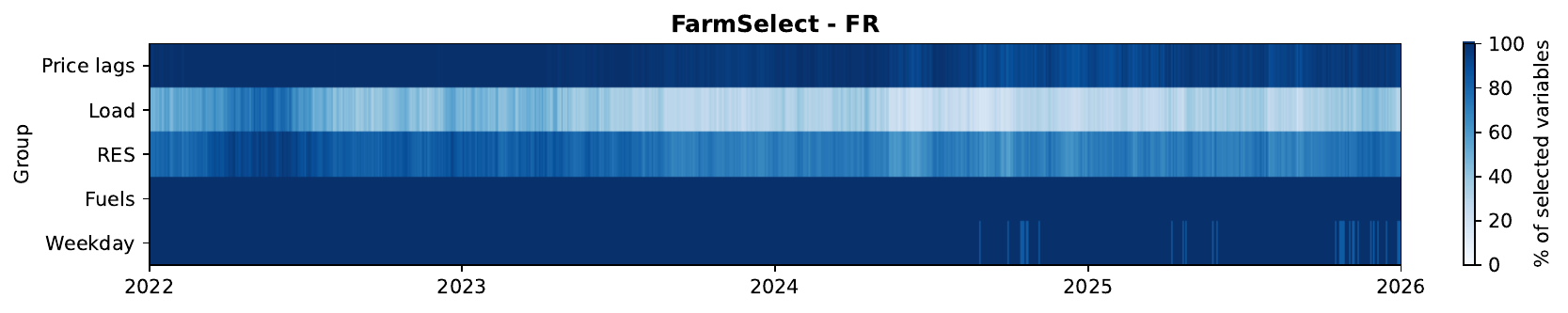}
    \includegraphics[width=\linewidth]{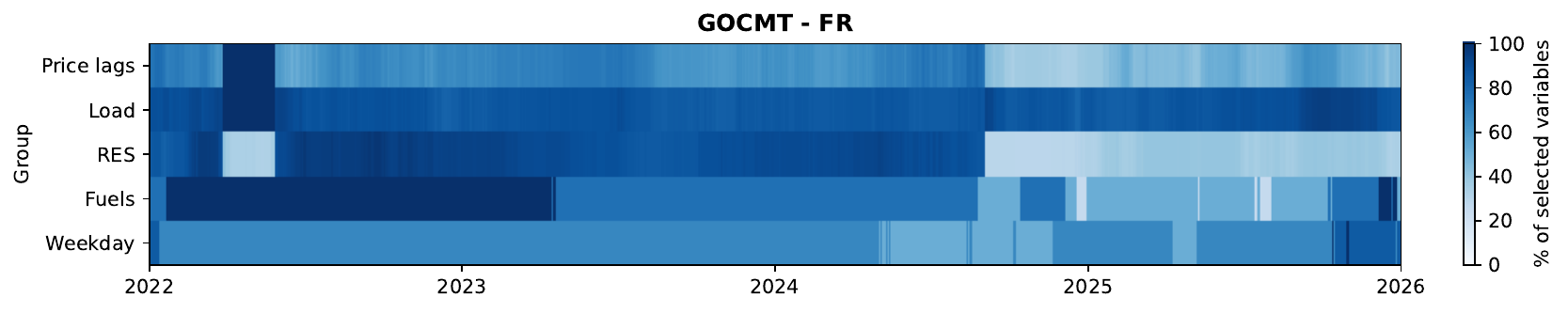}
    \caption{Heat map of the percentage of selected number of variables for each method from the groups Price lags, Load, RES, Fuels, and Weekday.}
\end{figure}

\begin{figure}[!ht]
    \centering
    \includegraphics[width=\linewidth]{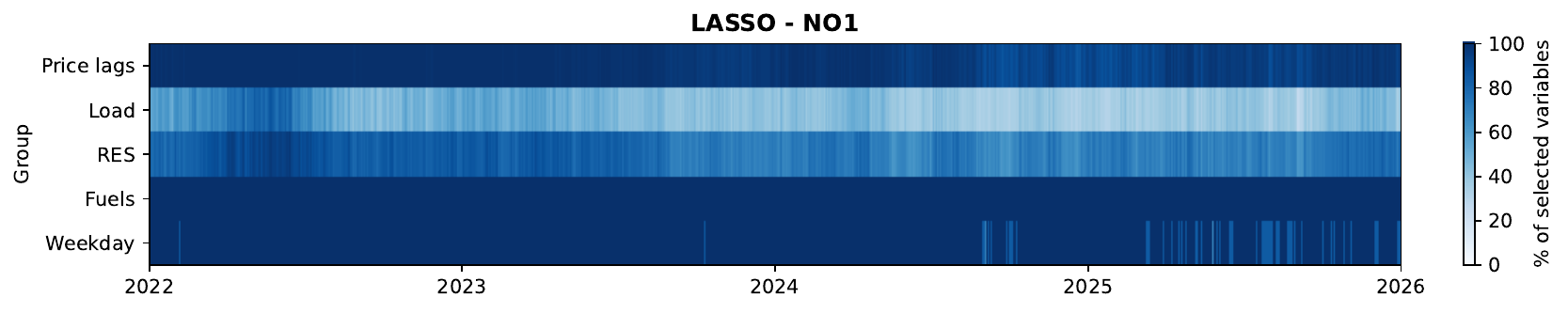}
    \includegraphics[width=\linewidth]{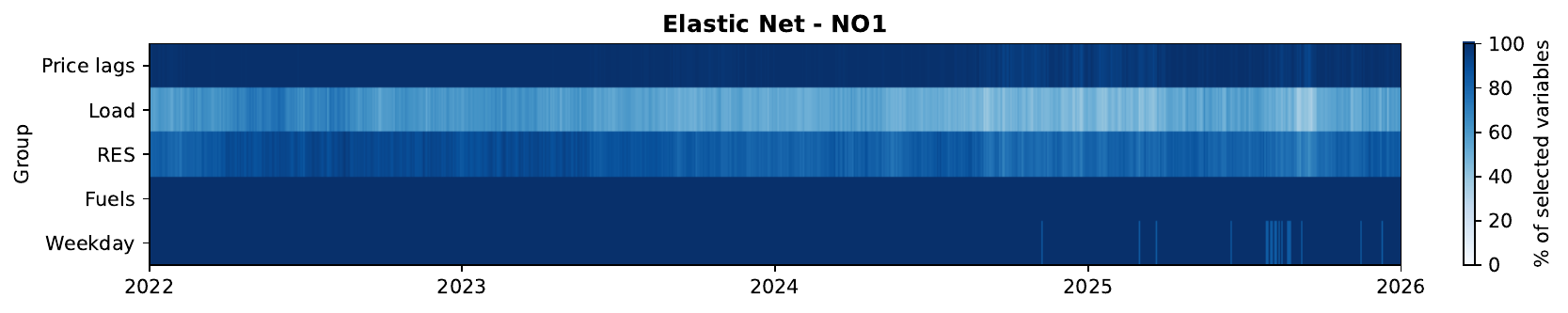}
    \includegraphics[width=\linewidth]{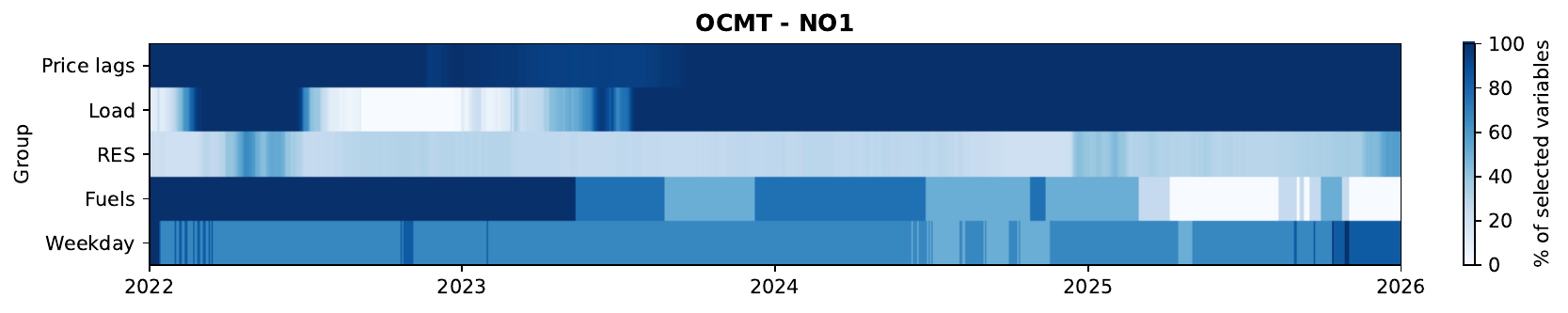}
    \includegraphics[width=\linewidth]{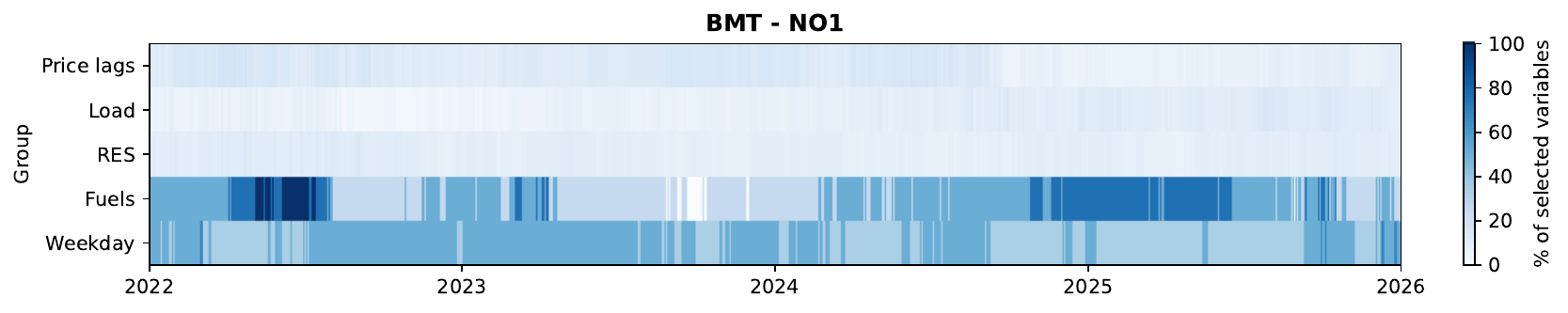}
    \includegraphics[width=\linewidth]{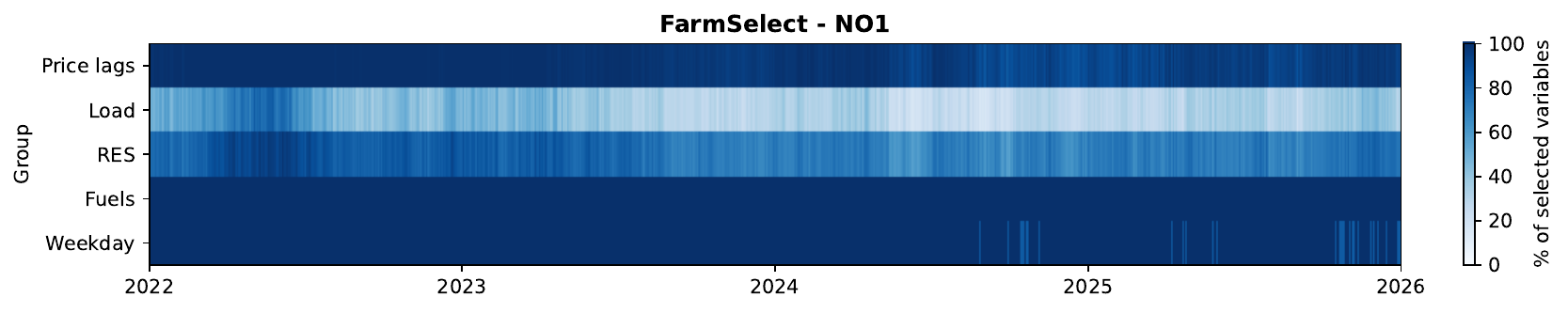}
    \includegraphics[width=\linewidth]{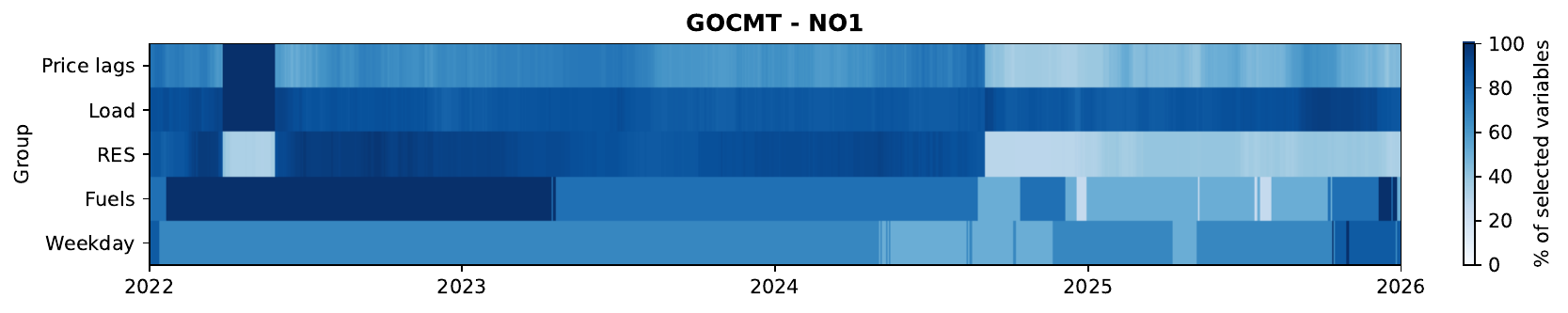}
    \caption{Heat map of the percentage of selected number of variables for each method from the groups Price lags, Load, RES, Fuels, and Weekday.}
\end{figure}

\begin{figure}[!ht]
    \centering
    \includegraphics[width=\linewidth]{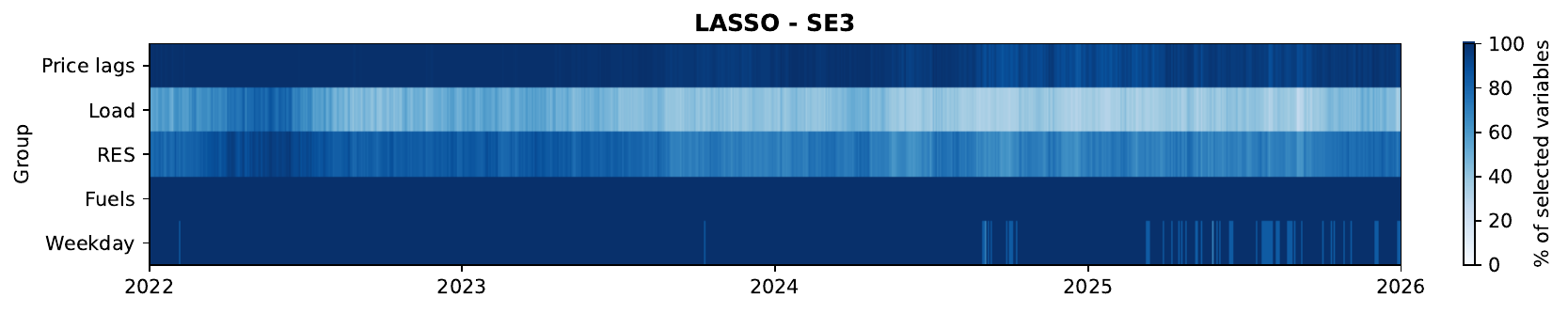}
    \includegraphics[width=\linewidth]{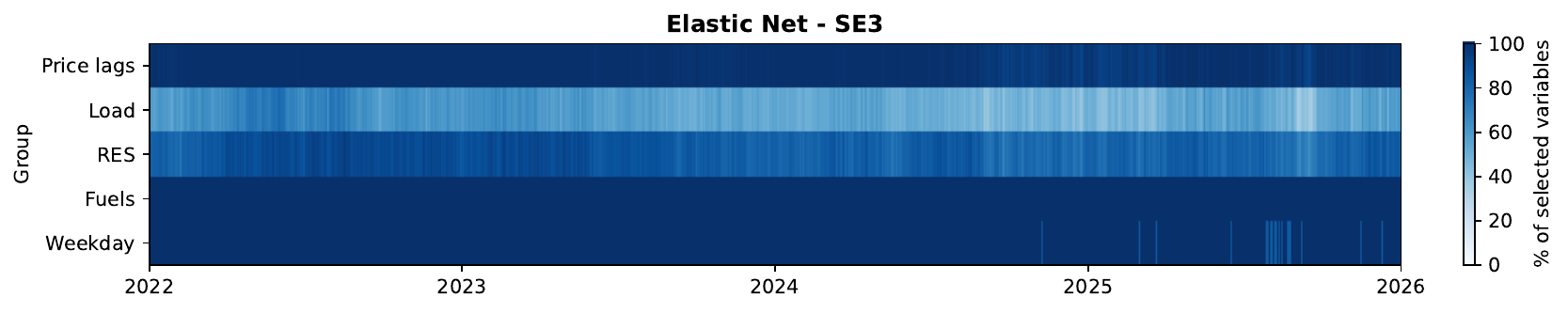}
    \includegraphics[width=\linewidth]{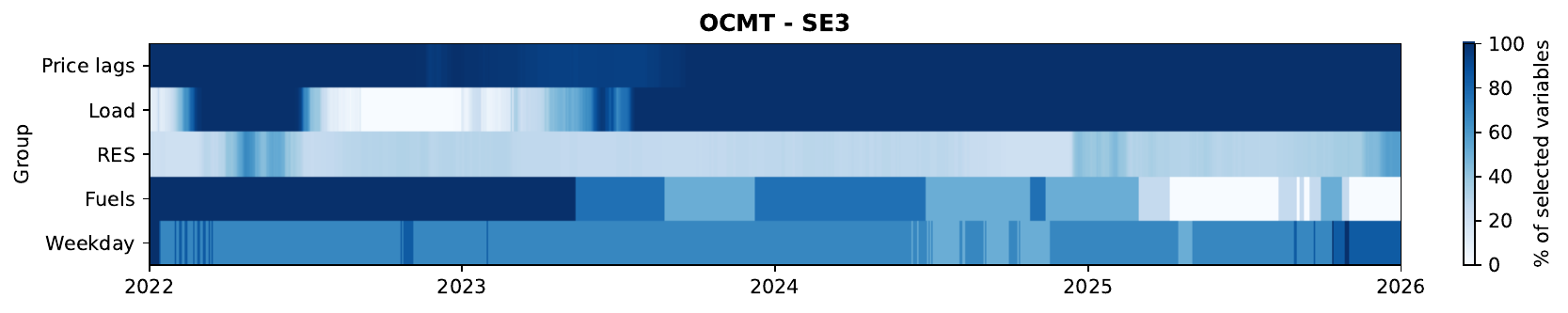}
    \includegraphics[width=\linewidth]{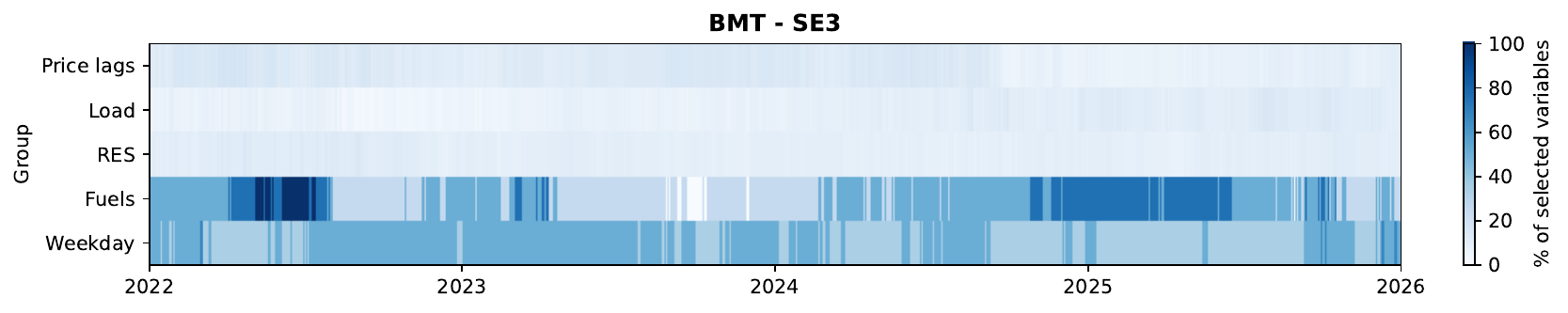}
    \includegraphics[width=\linewidth]{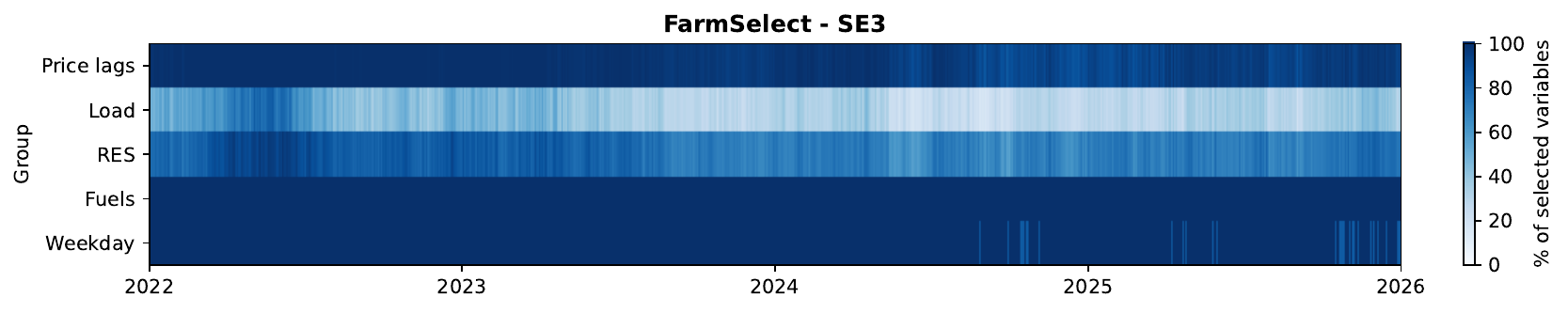}
    \includegraphics[width=\linewidth]{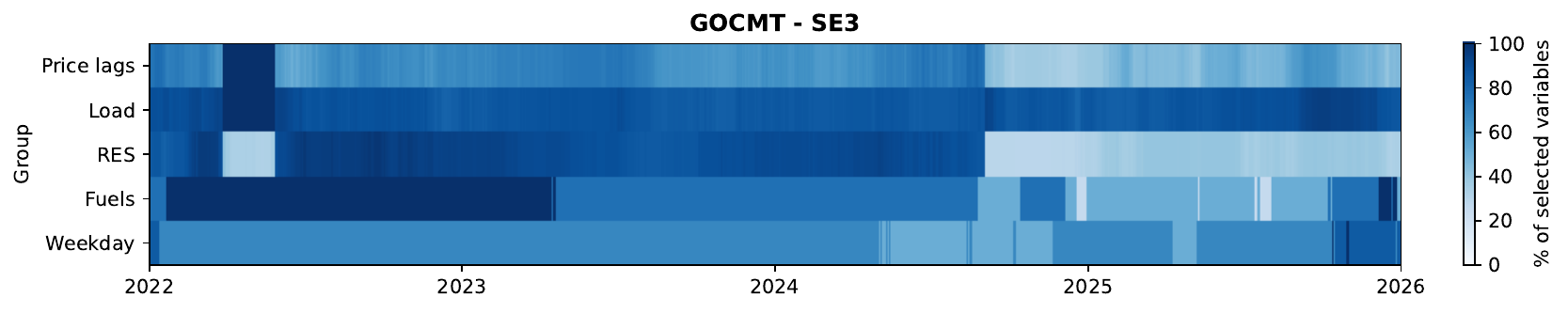}
    \caption{Heat map of the percentage of selected number of variables for each method from the groups Price lags, Load, RES, Fuels, and Weekday.}
\end{figure}

\end{document}